\documentclass[aps,prd,superscriptaddress,nofootinbib,floatfix,twocolumn]{revtex4}
\usepackage[dvipdfmx]{graphicx} 
\usepackage{mediabb}
\usepackage{amsmath}
\usepackage{bm}
\usepackage[small,bf]{caption}
\usepackage[subrefformat=parens]{subcaption}
\newcommand{\beq}{\begin{equation}}
\newcommand{\eeq}{\end{equation}}
\newcommand{\beqa}{\begin{eqnarray}}
\newcommand{\eeqa}{\end{eqnarray}}

\newcommand{\MP}{M_{\rm P}}
\newcommand{\nnb}{\nonumber}

\newcommand{\om}{\omega}
\newcommand{\vth}{\vartheta}

\newcommand{\mcal}[1]{\mathcal{#1}}
\newcommand{\mrm}[1]{\mathrm{#1}}
\newcommand{\diff}[2]{\frac{d#1}{d#2}}
\newcommand{\pdiff}[2]{\frac{\partial#1}{\partial#2}}
\newcommand{\Int}[2]{\int_{#1}^{#2}}

\newcommand{\B}{\langle}
\newcommand{\K}{\rangle}

\begin{document}

\title{
Stimulated radar collider for testing a model of dark energy
}

\author{Kensuke Homma}
\email[corresponding author: ]{khomma@hiroshima-u.ac.jp}
\author{Yuri Kirita}
\affiliation{Graduate School of Science, Hiroshima University, Kagamiyama, Higashi-Hiroshima, Hiroshima 739-8526, Japan} 

\date{\today}

\begin{abstract}
We propose a stimulated pulsed-radar collider 
for testing a dilaton model with the mass of $\sim 10^{-7}$ eV as a candidate of dark energy.
We have extended formulae for stimulated resonant photon-photon scattering 
in a quasi-parallel collision system by including fully asymmetric collision cases. 
With a pulse energy of 100 J in the GHz-band, for instance, 
which is already achieved by an existing klystron, 
we expect that the sensitivity can reach gravitationally weak coupling domains,
if two key technological issues are resolved: pulse compression in time reaching 
the Fourier transform limit, and single-photon counting for GHz-band photons. 
Such testing might extend the present horizon of particle physics.
\end{abstract}

\maketitle

\section{Introduction}
Since Rutherford's experiment, the observation of quantum scattering processes caused 
by colliding energetic charged particles has unveiled deeper layers of nature at the microscopic. 
With knowledge gleaned from these particle collisions, the Standard Model (SM) of elementary particles 
is now almost confirmed, with the recent discovery of the Higgs boson 
providing another point of evidence for the SM. However, the SM is still unsatisfactory 
when trying to quantitatively understand the profile of the energy density of the universe 
as evaluated from macroscopic gravitational observables through curvatures in spacetime. 
In particular, the SM can explain only $\sim 5$\% of the observed energy density of the universe. 
The remainder of the energy density is assumed to be accounted for by dark matter and dark energy~\cite{PDG}.
Why, then, is so little understood about dark components ? 

The gravitational coupling strength 
$G_N \sim 10^{-38}$~GeV${}^{-2}$ ($\hbar = c = 1$) is extraordinary weak, 
even relative to the weakest coupling strength of 
the weak interaction $G_F \sim 10^{-5}$~GeV${}^{-2}$ among the SM. 
Because of its extraordinary weakness, gravitational coupling has never been probed by elementary scattering processes.
In this sense, gravity has been, in practice, beyond the scope of experimental particle physics to date.
Therefore, it is unlikely that present knowledge of particle physics is 
sufficient to understand dark components obtained from gravitational observables.
We suggest in this paper that we can actually test scattering processes,
even with gravitationally weak coupling, if a properly designed stimulated photon--photon collider
is used. Such testing might extend the present horizon of particle physics.

Massless Nambu--Goldstone fields accompany spontaneous breaking of 
global continuous symmetries~\cite{NGB}.
The neutral pion is a typical Nambu--Goldstone Boson (NGB). 
However, the physical mass is slightly greater than zero. This pseudo-NGB (pNGB) state is caused 
by chiral symmetry breaking in quantum chromodynamics (QCD). 
However, pNGBs are not limited to chiral symmetry.
In general, whenever a global symmetry of any type is broken, we may expect a pNGB to exist.
This viewpoint can be used as a robust guiding principle to search for something very low 
in mass in the Universe, even without knowing the details of individual dynamics.

In this paper, we focus on a dilaton field~\cite{Dilaton} as a pNGB caused by
breaking of dilatation (scale) symmetry for which 
fifth-force searches have been stimulated~\cite{FifthForce}.
This dilaton field can be a testable source of dark energy in laboratory experiments.
The discovery of an accelerating universe revived today's version of the cosmological constant, 
leaving fine-tuning and coincidence problems unresolved.  
The simplest known approach to explain these problems is to introduce a scalar field, the dilaton.
Fujii proposed a scalar-tensor theory with a cosmological constant $\Lambda$ (STTL)~\cite{STTL}
based on Jordan's scalar-tensor theory (STT) \cite{Jordan}, one of the best-known
alternatives to Einstein's General Relativity.
STTL yields the scenario of a decaying cosmological constant in the Einstein frame 
corresponding to the observational frame
resulting in $\Lambda_{\rm obs}\sim t^{-2}$, where the present age of the
universe, $t_0 =1.37 \times 10^{10}{\rm y}$, is re-expressed as $\sim
10^{60}$ in the reduced Planckian units with $c=\hbar =\MP (=(8\pi
G)^{-1/2} \sim 10^{18}~{\rm GeV})=1$.   Given the unification-oriented
expectation   $\Lambda \sim 1$ in these units, the decay behavior
provides us a natural way of understanding why the observed value may be
as small as $10^{-120}$. 
As of this writing, the most up-to-date evaluation of the gravitationally coupling dilaton mass 
in STTL is $(0.15 \sim 0.59) \times 10^{-6}$ eV, based on the self-energy correction 
via Higgs two-loop diagrams~\cite{DilatonMass}.

To directly probe this low-mass dilaton field, photon--photon scattering
has special advantages arising from the coupling of two photons, 
because photons are massless and the center-of-mass system (cms) energy, $E_{cms}$, 
can be extremely low in comparison with that of charged particle collisions.
We therefore discuss the following effective Lagrangian, which expresses coupling to two photons:
\beq\label{eq1}
-{\cal L} = gM^{-1} \frac{1}{4}F_{\mu\nu}F^{\mu\nu} \phi,
\eeq
where a scalar-type field $\phi$ with effective coupling $g/M$ to two photons is assumed.
In the case of the dilaton, the predicted parameter space has $M=\MP$ and $g = (1/2 \sim 5)/(3\pi)\alpha_{qed}$~\cite{DilatonMass}
with $\alpha_{qed} = 1/137$, implying that experiments
are required to be sensitive to the gravitationally weak coupling domain.
\begin{figure}
 \begin{minipage}[t]{0.99\linewidth}
  \centering
  \subcaption{}\label{Fig1a}
  \includegraphics[keepaspectratio, scale=0.30]
  {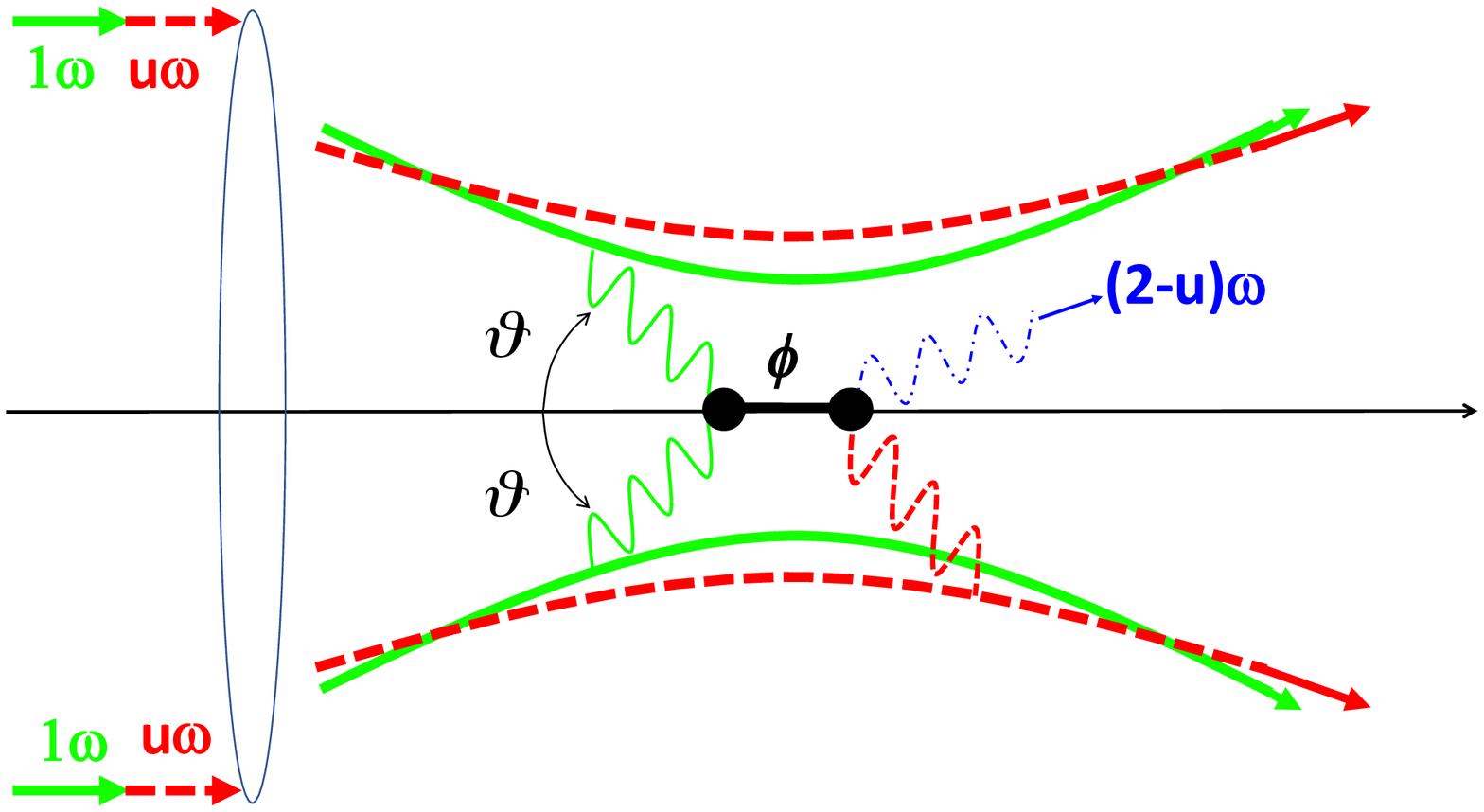}
 \end{minipage} \\
\begin{tabular}{cc}
 \begin{minipage}[b]{0.49\linewidth}
  \centering
  \subcaption{}\label{Fig1b}
  \includegraphics[keepaspectratio, scale=0.20]
  {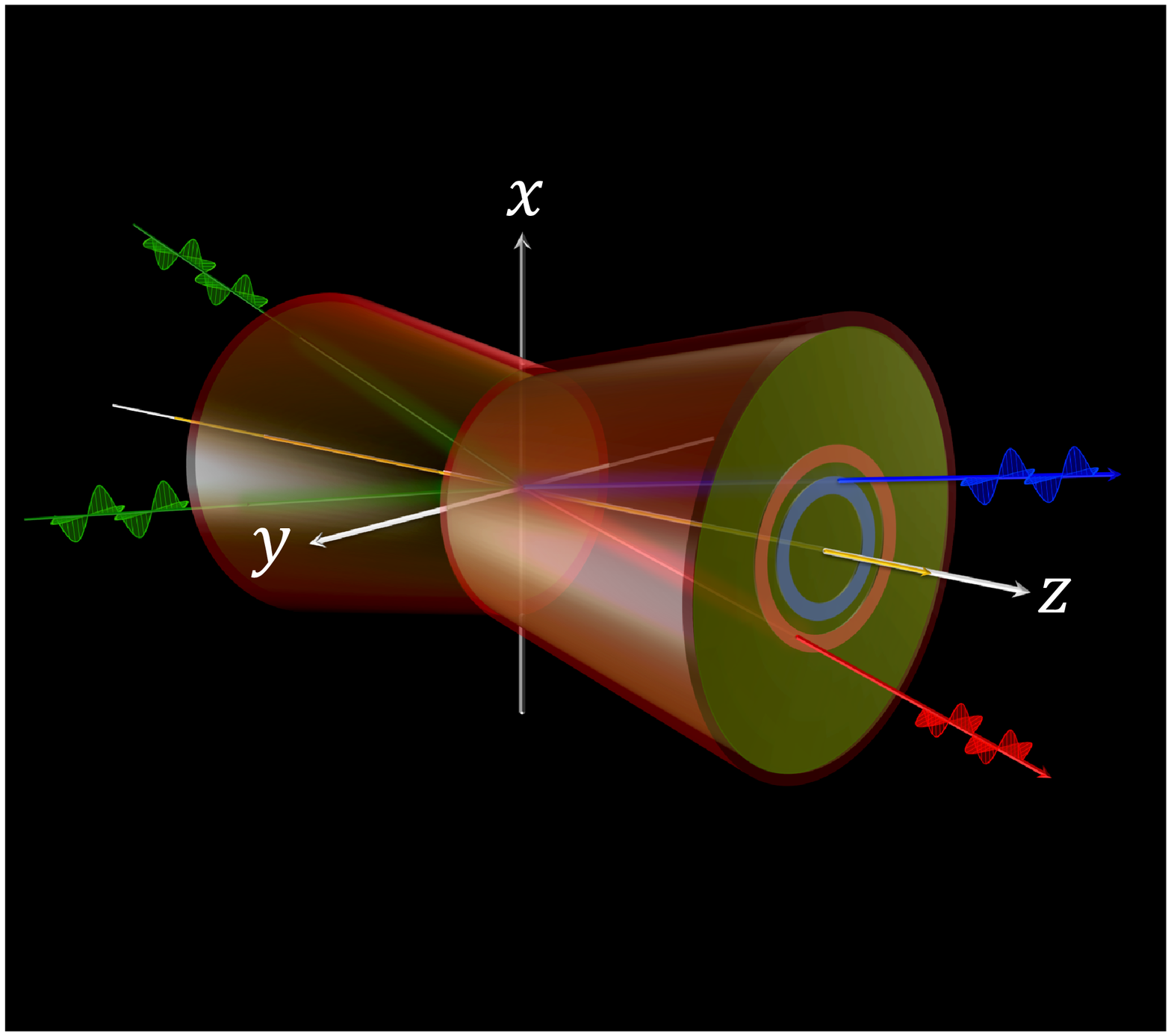}
 \end{minipage}
&
 \begin{minipage}[b]{0.49\linewidth}
  \centering
  \subcaption{}\label{Fig1c}
  \includegraphics[keepaspectratio, scale=0.20]
  {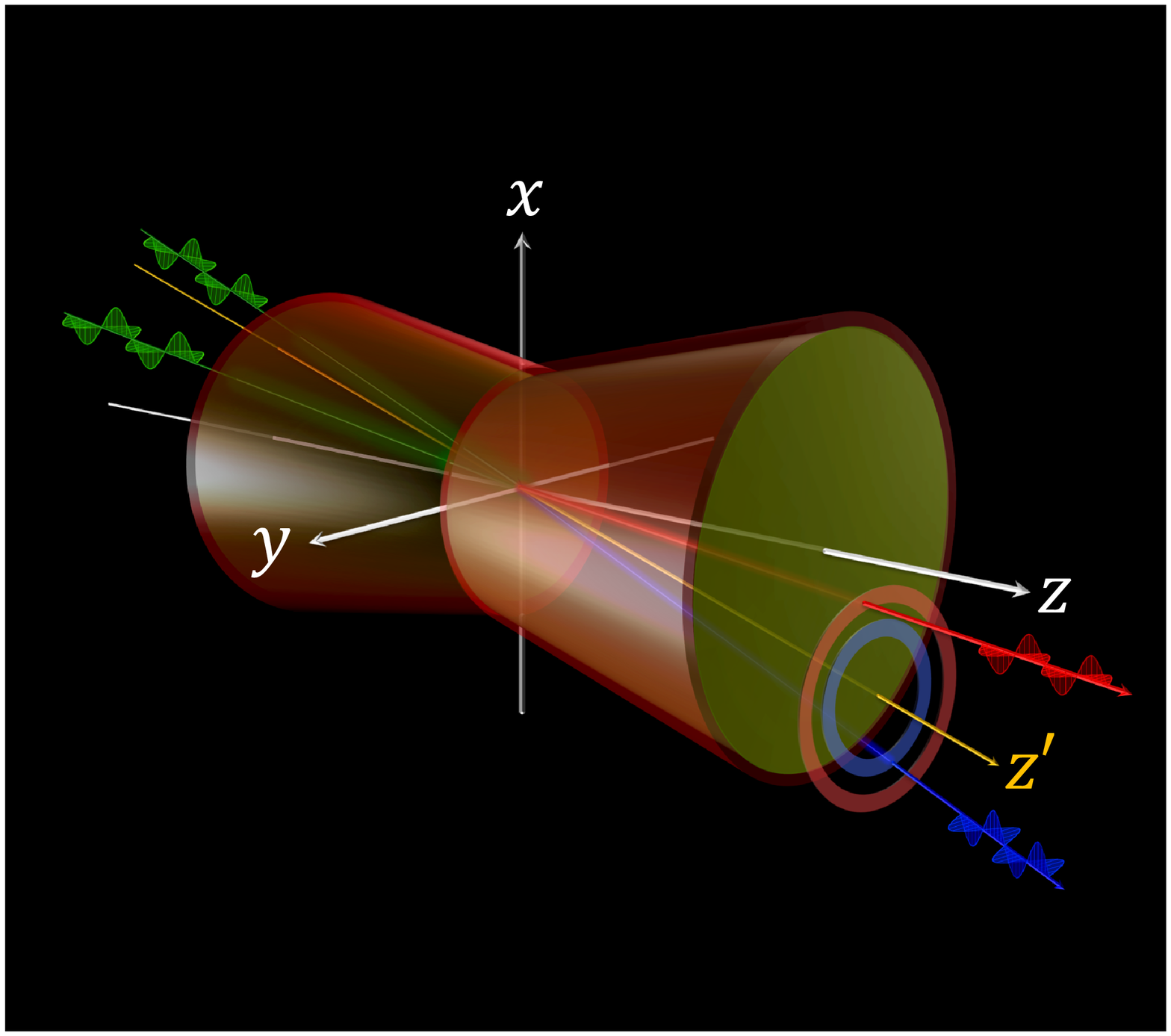}
 \end{minipage}
\end{tabular}
\caption{
Stimulated resonant scattering in a quasi-parallel collision system (QPS)
and the classification of collision geometries.
(a) Conceptual drawing of stimulated resonant photon--photon scattering in QPS, realizable
when a coherent field with energy $\omega$ (solid green line) is combined with another coherent field 
with energy $u\omega~(0<u<1)$ (dashed red line) and both fields are focused by a lens element
in vacuum. The emission of signal photons with energy $(2-u)\omega$ (dash-dotted blue line) is 
stimulated as a result of energy-momentum conservation in the scattering process
$\omega + \omega \rightarrow \phi \rightarrow (2-u)\omega + u\omega$ via a resonance
state $\phi$.
(b) Symmetric-incident and coaxial scattering, where the incident angles of two photon wave vectors
and their energies are symmetric, and the transverse momenta of photon pairs, $p_T$, 
always vanish with respect to the common optical axis $z$.
(c) Asymmetric-incident and non-coaxial scattering, where the incident angles of two photon wave
vectors and their energies are asymmetric, resulting in a finite value of
$p_T$ with respect to the common optical axis $z$. The zero-$p_T$ axis ($z^{'}$-axis)
is always configurable for arbitrary pairs of two incident wave vectors. 
}
\label{Fig1}
\end{figure}

We have advocated that stimulated photon--photon scattering 
in a Quasi-Parallel collision System (QPS),
illustrated in Fig.\ref{Fig1a}, can drastically enhance the interaction rate~\cite{PTP-DE}.
Capturing a resonance state in an $s$-channel photon--photon scattering
within the uncertainty on $E_{cms}$ is the first key
element of the proposed method. The second key element is the enhancement of the
interaction rate by the technique discussed in Appendix, which relies on the stimulated
nature of the two-body photon--photon scattering process, adding a coherently co-propagating field
as the inducing field.
Among several possible collision geometries~\cite{PTP-DE,PTEP-EXP00,PTEP-EXP01,PTEP17},
QPS is the optimum geometry for the low mass range,
having the widest accessible mass range possible for a single collision geometry.
For simplicity, we have initially considered QPS with a symmetric incident angle $\vartheta$, 
as shown in Fig.\ref{Fig1a}~\cite{PTP-DE}. 
This can be realized by focusing a photon beam with a single photon energy $\omega$.
In this case, $E_{cms}$ is expressed as
\beq
E_{cms} = 2\omega\sin\vartheta.
\eeq
This allows experiments to have two knobs to handle $E_{cms}$.
The choice of combination between photon energies and incident angles depends on the trade-off
between the beam and sensor technologies. 
In QPS, description of 
the interaction is non-trivial due to the inherently wave-like nature of photons~\cite{PTEP-EXP00}.
As we show in detail in Appendix, the interaction rate is increased when electromagnetic waves are 
confined to a short time scale. If the waves are confined to a short duration,
then an energy uncertainty $\delta \omega$ must be introduced according to the uncertainty principle
for the energy-time relation or, equivalently, as a result of the Fourier transform from the time domain 
to the frequency domain.
In addition, around the focal spot, the momentum uncertainty is also maximized due to the spatial 
localization of a beam field again based on the uncertainty principle for the momentum-space relation. 
This implies that the incident angles of electromagnetic waves must also fluctuate strongly. 
These situations require us to depart from the simplest geometry (i.e., from assuming symmetric energies 
and symmetric incident angles) and use a fully asymmetric geometry in QPS, as illustrated in Fig.\ref{Fig1c}.
The extended parametrization associated with the fully asymmetric case is non-trivial,
and we will show this in Appendix.
The main finding allowed by the extension to the fully asymmetric case is that 
the probability that non-coaxial collisions (Fig.\ref{Fig1c})
will occur dominates the probability of coaxial collisions (Fig.\ref{Fig1b}).

In the following sections, we first consider the concept of the stimulated pulsed-radar collider.
We then evaluate the expected sensitivity based on the parametrization including the fully asymmetric 
collision cases in QPS. In order to reach the gravitationally coupling dilaton field,
we discuss two technological requirements toward the future laboratory search for the dilaton field.
Finally, our conclusion is given.

\begin{figure}[h]
\centering
\includegraphics[keepaspectratio, scale=0.40]{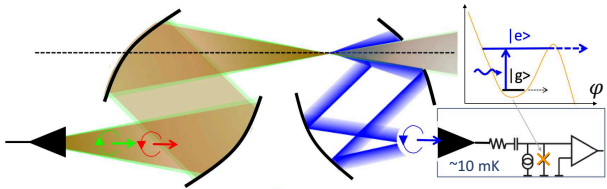}
\caption{Conceptual setup of the simulated radar collider and the detection of signal photons.
Two circularly polarized coherent beams for creation (green, left-handed) and 
stimulation (red, right-handed) are combined and focused along a common optical axis. 
Signal photons (blue, right-handed) are emitted via the exchange of a dilaton field.
Around the focal plane, only signal photons are partially reflected
and collimated by a dichroic parabolic mirror with a hole through which intense GHz beams can escape 
the detection system, both to avoid adding thermal background sources and to avoid picking up
atomic four-wave mixing processes from the upstream mirror surfaces as well as from residual gases
in the focal spot, because these background photons are expected to be confined 
within the incident angles of the two beams.
These peripherally emitted signal photons are focused into the detector element.
The detector consists of a signal photon counter. For sensing GHz photons, for instance,
a reasonable candidate is a Josephson-junction sensor based on a pulse-current-biased phase-qubit~\cite{Qbit1}. 
The bias instantaneously forms a potential, illustrated in the inset, 
as a function of the phase difference $\varphi$ between two superconductors sandwiching an insulator gap.
When a GHz-photon is absorbed in one of the two superconductors, the energy state of
a Cooper pair transits from the ground state $|g\K$ to an excited state $|e\K$,
which drastically increases the probability for the Cooper pair to tunnel to 
the neighboring superconductor though the gap. 
This allows number-resolved counting if parallelized junctions are implemented~\cite{Qbit1}.
}
\label{Fig2}
\end{figure}

\begin{figure*}[ht]
 \begin{minipage}[b]{0.49\linewidth}
  \centering
  \subcaption{}\label{Fig3a}
  \includegraphics[keepaspectratio, scale=0.41]
  {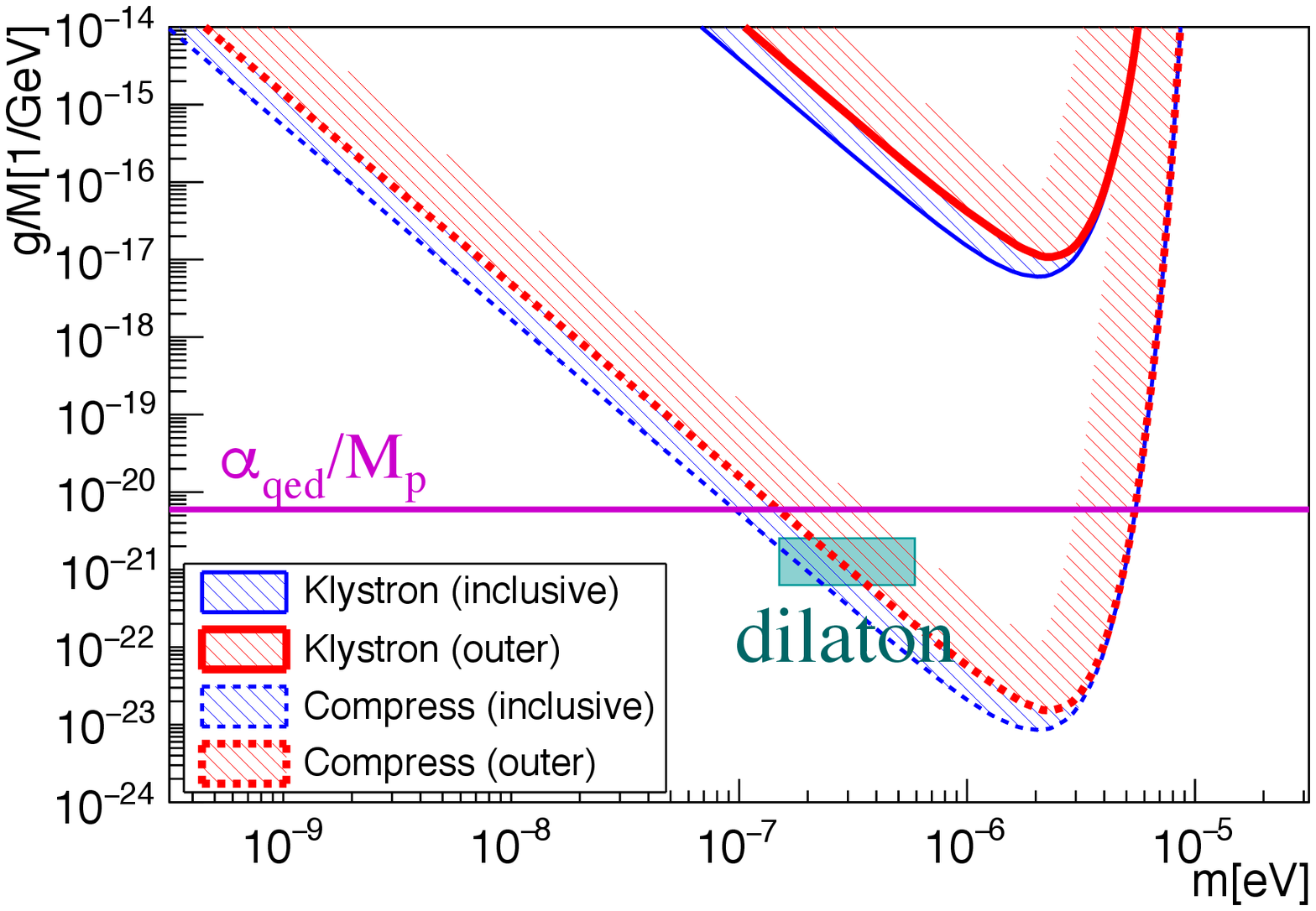}
 \end{minipage}
 \begin{minipage}[b]{0.49\linewidth}
  \centering
  \subcaption{}\label{Fig3b}
  \includegraphics[keepaspectratio, scale=0.46]
  {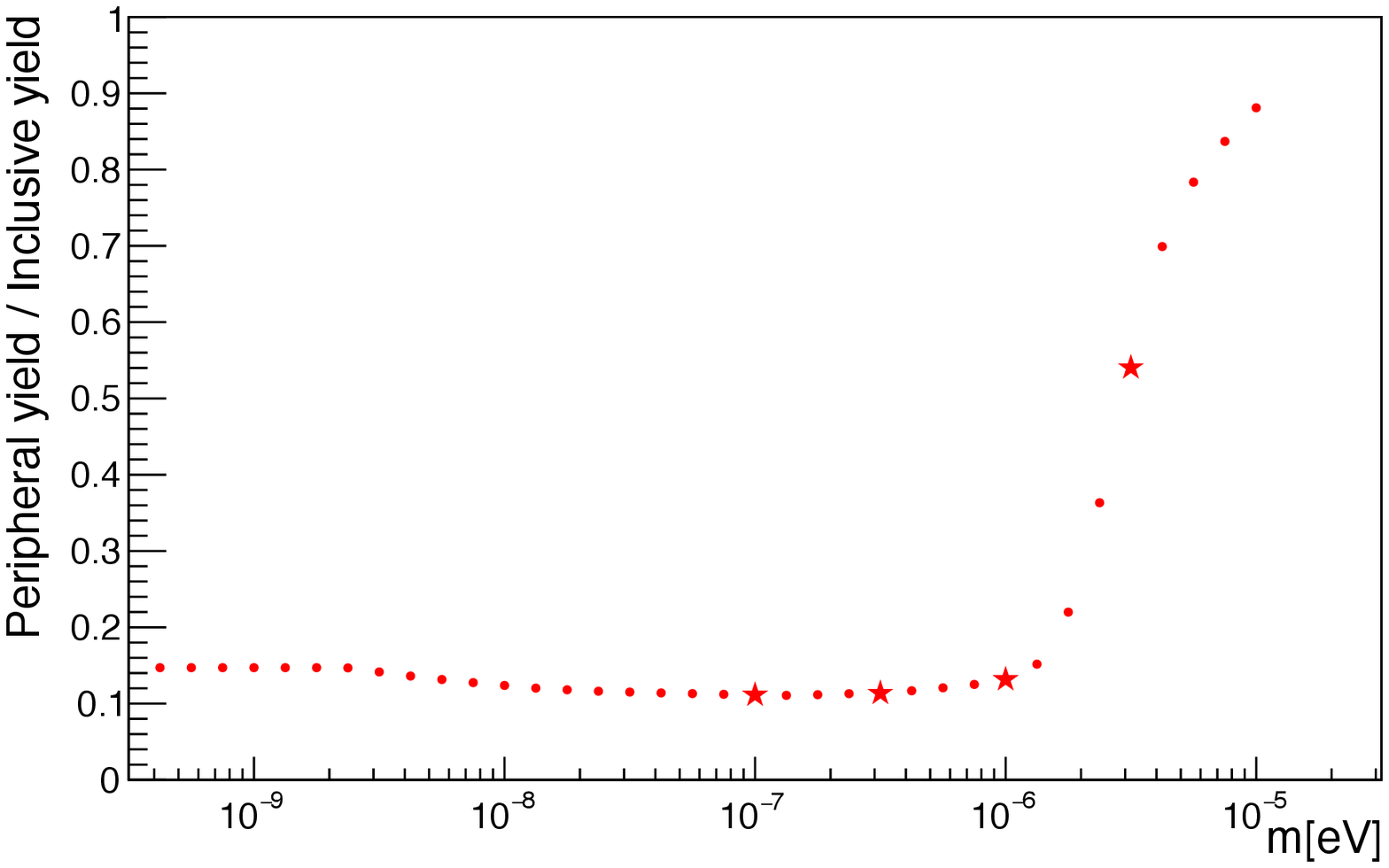}
 \end{minipage}
\\
 \begin{minipage}[b]{0.24\linewidth}
  \centering
  \includegraphics[keepaspectratio, scale=0.225]
  {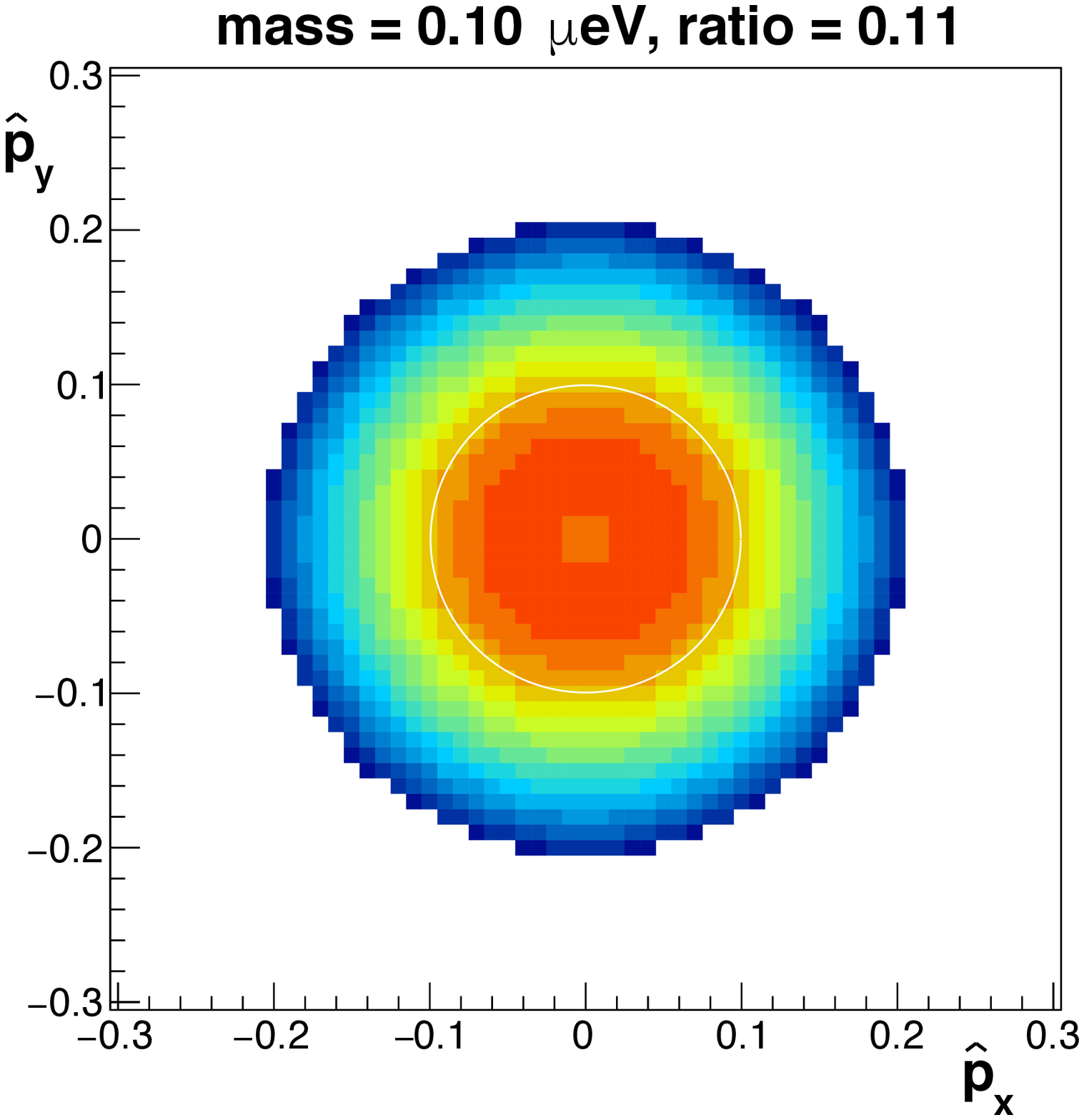}
 \end{minipage}
 \begin{minipage}[b]{0.24\linewidth}
  \centering
  \includegraphics[keepaspectratio, scale=0.225]
  {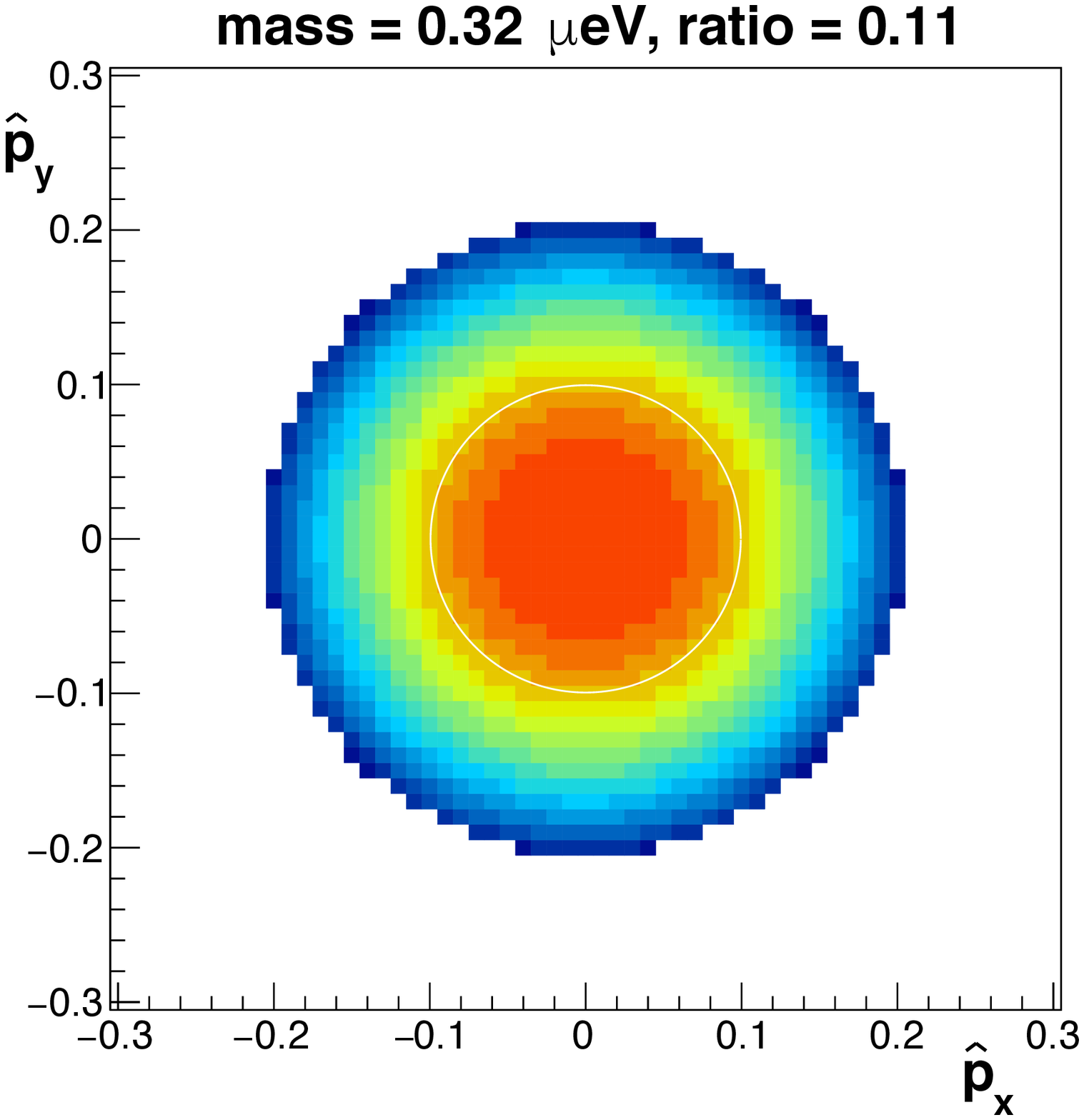}
 \end{minipage}
 \begin{minipage}[b]{0.24\linewidth}
  \centering
  \includegraphics[keepaspectratio, scale=0.225]
  {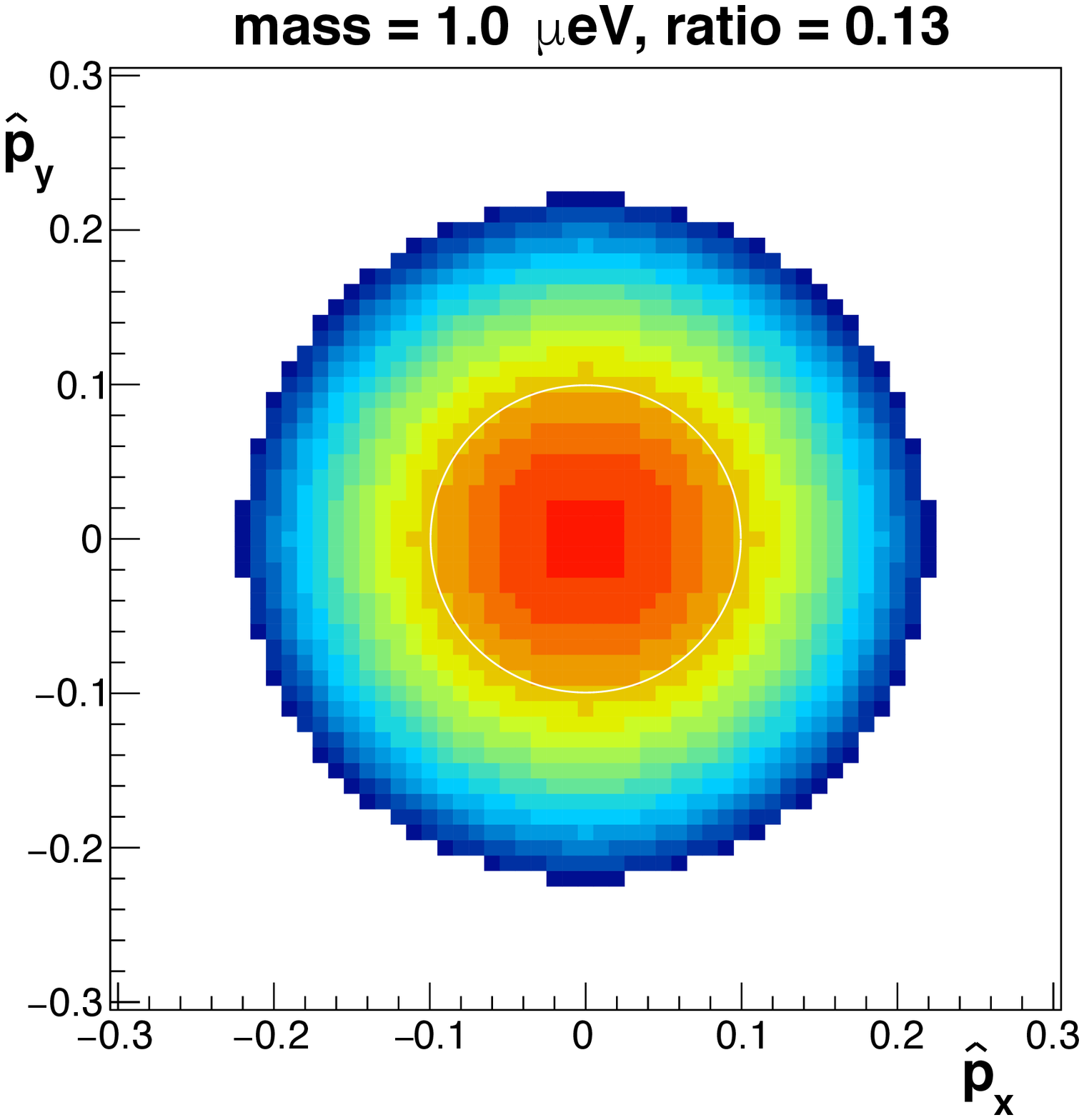}
 \end{minipage}
 \begin{minipage}[b]{0.24\linewidth}
  \centering
  \includegraphics[keepaspectratio, scale=0.225]
  {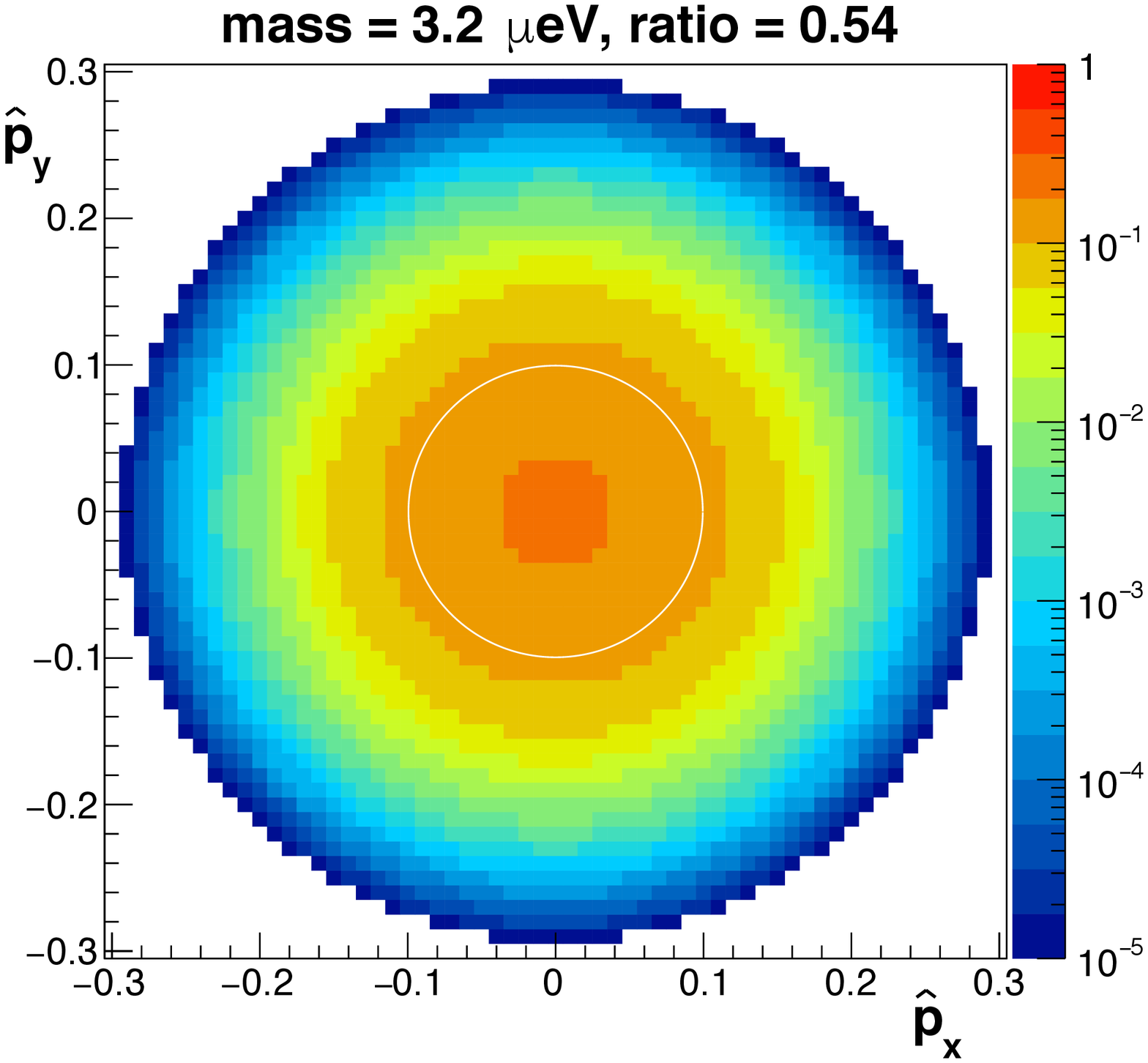}
 \end{minipage}
\caption{
Sensitivity curves and the ratio of outer emissions with the parameters listed in Tab.\ref{Tab1}. 
(a) Reachable coupling $g/M$ vs. dilaton mass $m$ domains.
The shaded rectangular domain corresponds to the prediction from the dilaton model~\cite{DilatonMass}.
The domains above the solid curves show the reachable range with peak power and 
relative linewidth of GHz-photon pulses equivalent to an existing 100~MW klystron~\cite{E3712}
(see the left column of Tab.\ref{Tab1} for the linewidth and pulse duration parameters). 
The dotted curves show the extended sensitivities when a relatively broader linewidth
(the right column for the same parameters in Tab.\ref{Tab1})
is used; it is assumed a system capable of this will be developed in the near future.
The thicker curves show the sensitivities when the signal photons are sampled 
only outside the divergence angles of the incident two beams.
(b) The ratio of the signal photon yield emitted outside the divergence angles of
incident beams to that of the inclusive yield as a function of mass. 
The lower four panels show the numerically calculated signal yields sampled at four mass points
indicated with the star markers in the ratio plot
as a function of the $x$ and $y$ components of unit momenta of
signal photons, $(\hat{p}_x, \hat{p}_y)$, at the beam waist $z=0$;
the white circles indicate the domains of beam divergence with the radius of $\sin\Theta_0$ from
Eq.(\ref{eq_Theta0}) in Appendix.
The color contours in the log scale are normalized to the common total number of signal photons.
}
\label{Fig3}
\end{figure*}

\section{Concept of stimulated pulsed-radar collider}
Toward the direct detection of the dilaton field of $m \sim 10^{-7}$~eV, 
if we consider only a laser source with single-photon energy $\sim{}1$~eV,
the incident angle must be maintained at $\sim{}10^{-7}$~radians and 
it is likely impossible to perform the search on the ground 
with a focal distance greater than $10^7$~m, assuming a beam diameter of $\sim{}1$~m. 
However, if we could use
an energy of $\omega \sim 10^{-5}$~eV with incident angle $\vartheta \sim 10^{-2}$~radians, it would be possible to
focus the beam to within a few hundred meters for a beam having a diameter of a few meters
and wavelength below 30~cm (1~GHz in frequency).
Indeed, intense sources of GHz photons are already commercially available, such 
as the klystron~\cite{E3712}.
Moreover, the number of photons per pulse is $\sim 10^5$ times 
that of optical laser fields for the same pulse energy.
Therefore, pulsed-radar beam in the GHz-band would be useful for a mass domain of $\sim 10^{-7}$~eV.
On the other hand, GHz photon counting with single-photon sensitivity is 
a difficult technological issue.
In the area of GHz-photon sensing, some successful examples of single-photon detection
with quantum-bit (qubit) technology have been reported recently ~\cite{Qbit1,Qbit2}.
In particular, a phase-qubit sensor based on Josephson junctions~\cite{Qbit1}
can be operated with a pulsed current bias within several ns duration~\cite{Qbit1}.
This time-gated operation would reduce dark currents from continuous 
background blackbody radiation.
Using the above considerations, Fig.\ref{Fig2} illustrates a conceptual setup 
for a dilaton search in QPS with GHz-photon sources 
equipped with a phase-qubit-type GHz-photon counter for the detection of signal photons.

\begin{table}[h]
\begin{center}
\begin{tabular}{lll}
\hline
Creation radar pulse & Klystron & Compress\\
\hline
central frequency $\nu_c$ & 2.8~GHz\\
energy per pulse & 100 J &\\
repetition rate $f$ & $50$~Hz &\\
relative linewidth $\delta \nu_c / \nu_c \times 100$ & $\pm 10^{-2}$~\% & $\pm 5$~\%  \\
half pulse duration, $\tau_c$ & $1~\mu$s & 0.568~ns \\
\hline
Inducing radar pulse & Klystron & Compress\\
\hline
central frequency $\nu_i$ & 1.6~GHz\\
energy per pulse & 100 J &\\
repetition rate $f$ & $50$~Hz &\\
relative linewidth $\delta \nu_i / \nu_i \times 100$ & $\pm 10^{-2}$~\% & $\pm 5$~\% \\
half pulse duration, $\tau_i$ & $1~\mu$s & 0.995~ns \\
\hline
Measurement requirements & Klystron & Compress\\
\hline
central signal frequency $\nu_s$ & 4 GHz &\\
\# of observed signals $N_{obs}$ & $100$~photons &\\
data acquisition time $t_{a}$ & 1 month ($2.6 \times 10^6$) &\\
detector current bias time $\tau_b$ &  1~ns &\\
detector efficiency $\epsilon_d$ & 0.1 &\\
overall efficiency $\epsilon \equiv \epsilon_d(\tau_b/2\tau_i)$ & $5 \times 10^{-5}$ & $5 \times 10^{-2}$\\
\end{tabular}
\begin{tabular}{lc}
\hline
Geometric parameters in QPS\\
\hline
creation beam diameter & 6.02~m \\
inducing beam diameter & 6.10~m  \\
common focal length  & 30~m (five Rayleigh length)\\
\hline
\end{tabular}
\end{center}
\caption{GHz-band pulsed beam parameters
similar to the existing klystron~\cite{E3712} and the compressed case reaching
the Fourier transform limit $\tau_j \times 2\pi\delta \nu_j \ge 1/2$ for $j$ = $c$ or $i$.
}
\label{Tab1}
\end{table}

\section{Expected sensitivity}
Given this stimulated radar collider setup 
with the experimental parameters set as listed in Tab.\ref{Tab1},
we discuss how we can reach the gravitational coupling domain in $g/M < \alpha_{qed}/\MP$,
based on new formulas including an asymmetric quasi-parallel collision geometry, 
explained in Appendix in detail. 
In symmetric-incident and coaxial scattering in QPS (Fig.\ref{Fig1b}),
transverse momenta of photon pairs, $p_T$, always vanish with respect to the common optical axis $z$.
This guarantees that azimuthal angles of the final state photon wave vectors are axially symmetric
around the $z$-axis. Therefore, the inducible momentum or angular range can be analytically obtained
via the axial symmetric nature of the focused beams.
On the other hand, in asymmetric-incident and non-coaxial scattering (Fig.\ref{Fig1c}),
finite transverse momenta are unavoidably introduced. 
However, a zero-$p_T$ axis, defined as the $z^{'}$-axis, is always 
configurable for any arbitrary pair of two incident wave vectors. Therefore, $z^{'}$-axis can restore 
the axial symmetric nature of the azimuthal angles of the final state wave vectors.
Despite this, the inducing coherent field is physically mapped to the common optical axis $z$.
Therefore, the inducible momentum range changes in a complicated manner that depends on an arbitrarily formed
$z^{'}$-axis. Hence, numerical integration must be performed 
to express the number of signal photons per shot ${\cal Y}_{c+i}$
in Eq.(\ref{eq_Yci}) by substituting Eqs.(\ref{eq_Sibar}) and (\ref{eq_Di}), shown in Appendix.
The number of experimentally observable signal photons $N_{obs}$ as a function of
mass and coupling for the set of experimental parameters $P$ given in Tab.\ref{Tab1} is then expressed as
\beq
N_{obs} = {\cal Y}_{c+i}(m, g/M ; P) t_{a} f \epsilon
\eeq
where the data acquisition time is $t_{a}$, the repetition rate of pulsed beams is $f$, and
the overall efficiency is $\epsilon \equiv \epsilon_d (\tau_b / 2\tau_{i})$
with detection efficiency $\epsilon_d$, 
qubit current-bias time $\tau_b$, and inducing pulse duration $\tau_i$.
By numerically solving this equation, we can obtain $g/M$ for the given values of $m$ and $N_{obs}$.
Dominant background photons are expected from blackbody radiations in
the same spectrum width as that of the signal photons, $(1 \pm \sim 0.05)\nu_s$. 
The unavoidable blackbody source is the entrance horn connected to the qubit senor. 
The total number of background photons is evaluated as $N_{bkg}=0.5$ photons
by assuming that the horn and sensor temperatures are 
kept at $T=10$~mK with the inner surface area of the cone-type horn $\Delta S = \pi\lambda^2_s$ 
for signal photon wavelength $\lambda_s=c/\nu_s$ and solid angle $\Delta \Omega=2\pi$
from the following relation
\beq
N_{bkg} = \Int{0.95\nu_s}{1.05\nu_s}
\frac{2h\nu^3}{c^2} \frac{1}{e^{h\nu/(kT)}-1} d\nu 
\frac{\Delta \Omega \Delta S t_{a} f \epsilon}{h \nu_s}.
\eeq
We also note that photon--photon scattering in the SM can be neglected
because the QED-based stimulated scattering is sufficiently suppressed 
by the $E^6_{cms}$ dependence of the cross section~\cite{PTEP-GG,PTEP17}.
Considering systematic backgrounds, we require $N_{obs} = 100 \gg N_{bkg}$ in this paper.
Figure \ref{Fig3a} summarizes the accessible domains for coupling $g/M$ versus mass $m$.
This indicates that broadening the linewidth is indeed a key factor 
because it can increase the interaction rate by increasing the spacetime
overlapping factor of the incident pulsed beams, as explained in Eq.(\ref{eq_Di}),
due to the short durations $\tau_c$ and $\tau_i$, 
and also increases the chance to stimulate emission of final
state photons satisfying energy-momentum conservation within the allowed energy-momentum fluctuations
of collision beams, as indicated in Eq.(\ref{eq_Sibar}) in Appendix. 
Figure \ref{Fig3b} shows the ratio of signal photons found outside the angular divergence
of the focused beams based on the geometric optics as a function of mass. 
For larger masses, larger fractions of signal photons are emitted to the outer angles.
The non-coaxial collisions allow signal photon emission
outside the divergence angles of the focused beams defined by geometric optics.
Thanks to this scattering behavior, we can expect that the ratio between the number of
signal photons and beam photons could be improved 
if we could measure only peripheral emissions around the common optical axes,
as illustrated in Fig.\ref{Fig2}.

\section{Conclusion}
We have formulated stimulated resonant photon--photon scattering 
in QPS including asymmetric-incident and non-coaxial collisions. 
From the stimulated pulsed-radar collider concept, we expect that the sensitivity 
can reach the domain in which the dilaton field (a candidate for dark energy) is predicted to exist,
assuming two key technological issues are resolved: pulse compression in time
reaching the Fourier transform limit, and single-photon counting for GHz-band photons.
These are possible in principle but technologically challenging in practice. 
It is worth being striving for them, however, 
because they would allow direct probing of gravitationally weak scattering processes 
in laboratory experiments, which has not been done in the history of science.

\begin{acknowledgments}
We express deep gratitude to Yasunori Fujii, who passed away in July 2019. 
This study is motivated by his outstanding works on the dilaton model and
the effort to evaluate the concrete mass of a dilaton as well as coupling.
We also thank K. Ishikawa for discussions about evaluation of the inducing effect,
S. Mima and S. Shibata for discussions on the qubit application, and 
M. Oxborrow for discussions on pulsed GHz sources.
K. Homma acknowledges the support of the Collaborative Research
Program of the Institute for Chemical Research,
Kyoto University (Grants Nos.\, 2018--83 and 2019--72)
and Grants-in-Aid for Scientific Research
Nos.\, 17H02897, 18H04354, and 19K21880 from the Ministry of Education, Culture, Sports, Science and Technology (MEXT) of Japan.
\end{acknowledgments}

\newpage
\section*{Appendix}
Here, we provide full details of the evaluation of signal yield in stimulated resonant photon--photon 
scattering in a quasi-parallel collision system (QPS) that includes fully asymmetric collision
and stimulation geometries due to uncertainties regarding energy and incident angles in QPS.
Figure \ref{Fig3} is calculated from numerical integration of Eq.(\ref{eq_Yci})
with Eqs.(\ref{eq_Sibar}) and (\ref{eq_Di}), using the parameter values given in Tab.\ref{Tab1}.
We use the metric convention $(+---)$ throughout this appendix.

\subsection{Lorentz-invariant transition amplitude in the sea of coherent fields}
The S-matrix for the interaction Lagrangian 
\beq
-{\cal L} = gM^{-1} \frac{1}{4}F_{\mu\nu}F^{\mu\nu} \phi
\eeq
is expressed as
\beqa\label{eq_Smatrix}
S^{(2)} = \left(-\frac{1}{4}\frac{g}{M}\right)^2 \frac{i^2}{2} \int d^4x \int d^4y \\ \nnb
\times T[F_{\mu\nu}(x)F^{\mu\nu}\phi(x)F_{\rho\sigma}(y)F^{\rho\sigma}(y)\phi(y)],
\eeqa
where $T$ denotes the time-ordered product.
From Wick's theorem, the T-product can be converted to the normal-ordering product
by requiring contractions with four external electromagnetic fields, as follows.
\beq\label{eq_NormalProduct}
N[F_{\mu\nu}(x)F^{\mu\nu}(x)F_{\sigma\rho}(y)F^{\sigma\rho}(y)
\B0|T[\phi(x)\phi(y)]|0\K]
\eeq
Here,
\beq\label{eq_Propagator}
i\B0|T[\phi(x)\phi(y)]|0\K \equiv 
\frac{1}{(2\pi)^4}\int d^4q \frac{e^{-iq(x-y)}}{m^2-q^2-i\epsilon}
\eeq
is the propagator of a massive scalar field $\phi$ with an infinitesimal number $\epsilon$.
We expand the field strength tensor as
\beq\label{eq_Fmn}
F^{\mu\nu} \equiv
(-i)\int \frac{d^3 \bm{p}}{(2\pi)^3 2p^0}\Sigma_{\lambda=1,2}
(P^{\mu\nu}e^{-ipx}a_{\bm{p},\lambda} + \hat{P}^{\mu\nu}e^{ipx}a^{\dagger}_{\bm{p},\lambda})
\eeq
and further define the following momentum-polarization tensors as capitalized symbols
for an arbitrary four-momentum $p$ of the electromagnetic field with the polarization state $\lambda$:
\beqa\label{eq_Tensor}
P^{\mu\nu} &\equiv& p^{\mu}\epsilon^{\nu}(p,\lambda)-\epsilon^{\mu}(p,\lambda)p^{\nu},
\\ \nnb
\hat{P}^{\mu\nu} &\equiv&  \epsilon^{*\mu}(p,\lambda)p^{\nu}-p^\mu\epsilon^{*\nu}(p,\lambda).
\eeqa
The commutation relations are
\beqa\label{eq_commutation}
[a_{\bm{k},\lambda},a^{\dagger}_{\bm{k}^{'},\lambda^{'}}] &=& 
(2\pi)^3 2p^0 \delta^3(\bm{k}-\bm{k}^{'})\delta(\lambda-\lambda^{'}), \\ \nnb
[a_{\bm{k},\lambda},a_{\bm{k}^{'},\lambda^{'}}] &=& [a^{\dagger}_{\bm{k},\lambda},a^{\dagger}_{\bm{k}^{'},\lambda^{'}}] = 0.
\eeqa
From here, we omit the polarization index $\lambda$ and 
the sum over it for the photon creation and annihilation operators,
$a_{\bm{p},\lambda}$ and $a^{\dagger}_{\bm{p},\lambda}$,
because we require fixed beam polarizations in the last step of the following calculations.
Substituting Eqs.(\ref{eq_NormalProduct}--\ref{eq_Tensor}) into (\ref{eq_Smatrix}),
we get
\beqa
S^{(2)} = \left(-\frac{1}{4}\frac{g}{M}\right)^2 \frac{i}{2} \int d^4x \int d^4y 
\int \frac{d^4q}{(2\pi)^4}\frac{e^{-iq(x-y)}}{m^2-q^2-i\epsilon} \times \quad \\ \nnb
(-i)^4\int \frac{d^3 \bm{s}}{(2\pi)^3 2s^0} 
      \int \frac{d^3 \bm{t}}{(2\pi)^3 2t^0}
      \int \frac{d^3 \bm{u}}{(2\pi)^3 2u^0}
      \int \frac{d^3 \bm{v}}{(2\pi)^3 2v^0} \times \\ \nnb
T[\mbox{\hspace{2mm}}
\bigl(
S_{\mu\nu}T^{\mu\nu}       e^{-i(s+t)x}a_{\bm{s}}a_{\bm{t}} +
S_{\mu\nu}\hat{T}^{\mu\nu} e^{-i(s-t)x}a_{\bm{s}}a^{\dagger}_{\bm{t}} +
\quad \\ \nnb
\hat{S}_{\rho\sigma}T^{\rho\sigma} e^{-i(t-s)x}a^{\dagger}_{\bm{s}} a_{\bm{t}} +
\mbox{\hspace{2mm}} 
\hat{S}_{\rho\sigma}\hat{T}^{\rho\sigma} e^{i(s+t)x}a^{\dagger}_{\bm{s}}a^{\dagger}_{\bm{t}}
\bigr) \times \quad \\ \nnb
\bigl(
U_{\mu\nu}V^{\mu\nu}       e^{-i(u+v)y}a_{\bm{u}}a_{\bm{v}} +
U_{\mu\nu}\hat{V}^{\mu\nu} e^{-i(u-v)y}a_{\bm{u}}a^{\dagger}_{\bm{v}} +
\\ \nnb
\hat{U}_{\rho\sigma}V^{\rho\sigma} e^{-i(v-u)y}a^{\dagger}_{\bm{u}} a_{\bm{v}} +
\mbox{\hspace{2mm}} 
\hat{U}_{\rho\sigma}\hat{V}^{\rho\sigma} e^{i(u+v)y}a^{\dagger}_{\bm{u}}a^{\dagger}_{\bm{v}}
\bigr)\mbox{\hspace{2mm}}
].
\eeqa
Since we focus on only two-body--two-body interactions, the relevant S-matrix 
(including two creation and two annihilation operators) is expressed as
\beqa\label{eq_TwoBodySmatrix}
S^{(2)}_{2 \rightarrow 2} = \left(-\frac{1}{4}\frac{g}{M}\right)^2 \frac{i}{2}(2\pi)^4
\times \mbox{\hspace{2cm}}\\ \nnb
\int \frac{d^3 \bm{s}}{(2\pi)^3 2s^0} 
\int \frac{d^3 \bm{t}}{(2\pi)^3 2t^0}
\int \frac{d^3 \bm{u}}{(2\pi)^3 2u^0}
\int \frac{d^3 \bm{v}}{(2\pi)^3 2v^0} \times \\ \nnb
\bigl(
G_{-s,-t} \delta^4(-u-v+s+t) S_{\mu\nu}T^{\mu\nu}\hat{U}_{\rho\sigma}\hat{V}^{\rho\sigma}
a^{\dagger}_{\bm{u}} a^{\dagger}_{\bm{v}} a_{\bm{s}}a_{\bm{t}}
+ \\ \nnb
G_{-s,+t} \delta^4(+u-v+s-t) S_{\mu\nu}\hat{T}^{\mu\nu}U_{\rho\sigma}\hat{V}^{\rho\sigma}
a^{\dagger}_{\bm{t}} a^{\dagger}_{\bm{v}} a_{\bm{s}}a_{\bm{u}}
+ \\ \nnb
G_{-s,+t} \delta^4(-u+v+s-t) S_{\mu\nu}\hat{T}^{\mu\nu}\hat{U}_{\rho\sigma}V^{\rho\sigma}
a^{\dagger}_{\bm{t}} a^{\dagger}_{\bm{u}} a_{\bm{s}}a_{\bm{v}}
+ \\ \nnb
G_{+s,-t} \delta^4(+u-v-s+t) \hat{S}_{\mu\nu}T^{\mu\nu}U_{\rho\sigma}\hat{V}^{\rho\sigma}
a^{\dagger}_{\bm{s}} a^{\dagger}_{\bm{v}} a_{\bm{t}}a_{\bm{u}}
+ \\ \nnb
G_{+s,-t} \delta^4(-u+v-s+t) \hat{S}_{\mu\nu}T^{\mu\nu}\hat{U}_{\rho\sigma}V^{\rho\sigma}
a^{\dagger}_{\bm{s}} a^{\dagger}_{\bm{u}} a_{\bm{t}}a_{\bm{v}}
+ \\ \nnb
G_{+s,+t} \delta^4(+u+v-s-t) \hat{S}_{\mu\nu}\hat{T}^{\mu\nu}U_{\rho\sigma}V^{\rho\sigma}
a^{\dagger}_{\bm{s}} a^{\dagger}_{\bm{t}} a_{\bm{u}}a_{\bm{v}}
\mbox{\hspace{2mm}}\bigr),
\eeqa
where
$G_{i, j} \equiv (m^2-(i + j)^2)^{-1}$ indicate corresponding propagators.

Let us recall the definition of the coherent state~\cite{Glauber}:
\beq
|N_{\bm{p}}\K\K \equiv \exp\left(-N_{\bm{p}}/2 \right) \sum_{n=0}^{\infty}
\frac{N^{n/2}_{\bm{p}}}{\sqrt{n!}} |n_{\bm{p}}\K,
\label{coh_2}
\eeq
where $|n_{\bm{p}}\K$ is the normalized state of $n$ photons
\beq
|n_{\bm{p}}\K =\frac{1}{\sqrt{n!}}\left( a^\dagger_{\bm{p}}\right)^n |0\K,
\label{coh_3a}
\eeq
with the creation operator $a^\dagger_{\bm{p}}$ 
of photons that share a common momentum $\bm{p}$ and a common polarization state
over different number states.
The following relations on the coherent state
\beq
\B\B N_{\bm{p}} | N_{\bm{p}} \K\K = 1
\eeq
and
\beq
\B\B N_{\bm{p}}|n|N_{\bm{p}}\K\K =
\B\B N_{\bm{p}}|\left(a^\dagger_{\bm{p}} a_{\bm{p}}\right)|N_{\bm{p}}\K\K=N_{\bm{p}},
\label{coh_3e}
\eeq
give us basic properties with respect to the creation and annihilation operators:
\beq\label{eq_enhancement}
a_{\bm{p}}|N_{\bm{p}}\K\K =\sqrt{N_{\bm{p}}}|N_{\bm{p}}\K\K ,\quad\mbox{and}\quad 
\B\B N_{\bm{p}}|a^\dagger_{\bm{p}} =\sqrt{N_{\bm{p}}}\B\B N_{\bm{p}}|.
\eeq
We first consider a search for signal photons $p_3$ via the scattering process 
$p_1 + p_2 \rightarrow p_3 + p_4$
by supplying coherent fields $|N_{\bm{p}_1}\K\K$, $|N_{\bm{p}_2}\K\K$ and $|N_{\bm{p}_4}\K\K$.
We then introduce initial and final states, respectively, as follows:
\beqa
|\Omega\K &\equiv& |N_{\bm{p}_1}\K\K |N_{\bm{p}_2}\K\K |N_{\bm{p}_4}\K\K |0\K, \mbox{and}\\ \nnb
\B \Omega^{'}| &\equiv& \B\B N_{\bm{p}_1}| \B\B N_{\bm{p}_2}| \B\B N_{\bm{p}_4}| \B 1_{\bm{p}_3}| 
= \B \Omega | a_{\bm{p}_3}.
\eeqa
The two-body transition amplitude 
$\B \Omega^{'} | S^{(2)}_{2\rightarrow2} | \Omega \K$
contains the common operator products 
$a^{\dagger}_{\bm{i}} a^{\dagger}_{\bm{j}} a_{\bm{k}} a_{\bm{l}}$.
We then separately evaluate the contractions with coherent bra- and ket-states, respectively, as follows:
\beqa
\B \Omega^{'} | 
a^{\dagger}_{\bm{i}} a^{\dagger}_{\bm{j}}
= \sqrt{N_{p_1}}\hat{\delta}^{3}(\bm{p}_1 - \bm{i}) \times \mbox{\hspace{3cm}}\\ \nnb
\{ 
\B \Omega^{'} | \sqrt{N_{p_1}}\hat{\delta}^{3}(\bm{p}_1 - \bm{j}) +
\B \Omega^{'} | \sqrt{N_{p_2}}\hat{\delta}^{3}(\bm{p}_2 - \bm{j}) + \mbox{\hspace{0cm}} \\ \nnb
\B \Omega^{'} | \sqrt{N_{p_4}}\hat{\delta}^{3}(\bm{p}_4 - \bm{j}) + 
\B \Omega     | 1             \hat{\delta}^{3}(\bm{p}_3 - \bm{j})
\} + \mbox{\hspace{0.7cm}} \\ \nnb
\sqrt{N_{p_2}}\hat{\delta}^{3}(\bm{p}_2 - \bm{i}) \times \mbox{\hspace{3cm}}\\ \nnb
\{ 
\B \Omega^{'} | \sqrt{N_{p_1}}\hat{\delta}^{3}(\bm{p}_1 - \bm{j}) + 
\B \Omega^{'} | \sqrt{N_{p_2}}\hat{\delta}^{3}(\bm{p}_2 - \bm{j}) + \mbox{\hspace{0cm}} \\ \nnb 
\B \Omega^{'} | \sqrt{N_{p_4}}\hat{\delta}^{3}(\bm{p}_4 - \bm{j}) +
\B \Omega     | 1             \hat{\delta}^{3}(\bm{p}_3 - \bm{j})
\} + \mbox{\hspace{0.7cm}} \\ \nnb
\sqrt{N_{p_4}}\hat{\delta}^{3}(\bm{p}_4 - \bm{i}) \times \mbox{\hspace{3cm}}\\ \nnb
\{ 
\B \Omega^{'} | \sqrt{N_{p_1}}\hat{\delta}^{3}(\bm{p}_1 - \bm{j}) + 
\B \Omega^{'} | \sqrt{N_{p_2}}\hat{\delta}^{3}(\bm{p}_2 - \bm{j}) + \mbox{\hspace{0cm}} \\ \nnb
\B \Omega^{'} | \sqrt{N_{p_4}}\hat{\delta}^{3}(\bm{p}_4 - \bm{j}) +
\B \Omega     | 1             \hat{\delta}^{3}(\bm{p}_3 - \bm{j})
\} + \mbox{\hspace{0.7cm}} \\ \nnb
1             \hat{\delta}^{3}(\bm{p}_3 - \bm{i}) \times \mbox{\hspace{3.5cm}}\\ \nnb
\{ 
\B \Omega | \sqrt{N_{p_1}}\hat{\delta}^{3}(\bm{p}_1 - \bm{j}) + 
\B \Omega | \sqrt{N_{p_2}}\hat{\delta}^{3}(\bm{p}_2 - \bm{j}) + \mbox{\hspace{0.2cm}}\\ \nnb
\B \Omega | \sqrt{N_{p_4}}\hat{\delta}^{3}(\bm{p}_4 - \bm{j}) +
\B \Omega | a^{\dagger}_{\bm{j}}
\}, \mbox{\hspace{2.5cm}}
\eeqa
where the last term vanishes because $\B 0| a^{\dagger}_{\bm{p}} = 0$, and
\beqa
a_{\bm{k}} a_{\bm{l}} | \Omega \K = 
\sqrt{N_{p_1}}\hat{\delta}^{3}(\bm{p}_1 - \bm{l})
\{ 
\sqrt{N_{p_1}}\hat{\delta}^{3}(\bm{p}_1 - \bm{k})|\Omega\K +  \mbox{\hspace{0.5cm}} \\ \nnb
\sqrt{N_{p_2}}\hat{\delta}^{3}(\bm{p}_2 - \bm{k})|\Omega\K +
\sqrt{N_{p_4}}\hat{\delta}^{3}(\bm{p}_4 - \bm{k})|\Omega\K
\} + \\ \nnb
\sqrt{N_{p_2}}\hat{\delta}^{3}(\bm{p}_2 - \bm{l})
\{ 
\sqrt{N_{p_1}}\hat{\delta}^{3}(\bm{p}_1 - \bm{k})|\Omega\K +  \mbox{\hspace{0.5cm}} \\ \nnb
\sqrt{N_{p_2}}\hat{\delta}^{3}(\bm{p}_2 - \bm{k})|\Omega\K +
\sqrt{N_{p_4}}\hat{\delta}^{3}(\bm{p}_4 - \bm{k})|\Omega\K
\} + \\ \nnb
\sqrt{N_{p_4}}\hat{\delta}^{3}(\bm{p}_4 - \bm{l})
\{ 
\sqrt{N_{p_1}}\hat{\delta}^{3}(\bm{p}_1 - \bm{k})|\Omega\K +  \mbox{\hspace{0.5cm}} \\ \nnb
\sqrt{N_{p_2}}\hat{\delta}^{3}(\bm{p}_2 - \bm{k})|\Omega\K +
\sqrt{N_{p_4}}\hat{\delta}^{3}(\bm{p}_4 - \bm{k})|\Omega\K
\}.
\eeqa
Because the search window is designed for the scattering process $p_1 + p_2 \rightarrow p_3 + p_4$,
the consistent transition amplitude that satisfies the combination of the initial and final state
momenta is limited to
\begin{widetext}
\beqa\label{AmpElement3}
\B \Omega^{'} | 
a^{\dagger}_{\bm{i}} a^{\dagger}_{\bm{j}} a_{\bm{k}}a_{\bm{l}}
| \Omega \K = 
\{ \sqrt{N_{p_4}}\hat{\delta}^{3}(\bm{p}_4 - \bm{i})1\hat{\delta}^{3}(\bm{p}_3 - \bm{j}) +
1\hat{\delta}^{3}(\bm{p}_3 - \bm{i}) \sqrt{N_{p_4}}\hat{\delta}^{3}(\bm{p}_4 - \bm{j}) \} 
\times \\ \nnb
\{ \sqrt{N_{p_1}}\hat{\delta}^{3}(\bm{p}_1 - \bm{l})\sqrt{N_{p_2}}\hat{\delta}^{3}(\bm{p}_2 - \bm{k}) + 
\sqrt{N_{p_2}}\hat{\delta}^{3}(\bm{p}_2 - \bm{l}) \sqrt{N_{p_1}}\hat{\delta}^{3}(\bm{p}_1 - \bm{k}) \} 
\B \Omega | \Omega \K \\ \nnb 
= \sqrt{N_{p_1}} \sqrt{N_{p_2}} \sqrt{N_{p_4}} 
\{
\hat{\delta}^{3}(\bm{p}_4 - \bm{i})\hat{\delta}^{3}(\bm{p}_3 - \bm{j})
\hat{\delta}^{3}(\bm{p}_2 - \bm{k})\hat{\delta}^{3}(\bm{p}_1 - \bm{l}) + \\ \nnb
\hat{\delta}^{3}(\bm{p}_4 - \bm{i})\hat{\delta}^{3}(\bm{p}_3 - \bm{j})
\hat{\delta}^{3}(\bm{p}_1 - \bm{k})\hat{\delta}^{3}(\bm{p}_2 - \bm{l}) + \\ \nnb
\hat{\delta}^{3}(\bm{p}_3 - \bm{i})\hat{\delta}^{3}(\bm{p}_4 - \bm{j})
\hat{\delta}^{3}(\bm{p}_2 - \bm{k})\hat{\delta}^{3}(\bm{p}_1 - \bm{l}) + \\ \nnb
\hat{\delta}^{3}(\bm{p}_3 - \bm{i})\hat{\delta}^{3}(\bm{p}_4 - \bm{j})
\hat{\delta}^{3}(\bm{p}_1 - \bm{k})\hat{\delta}^{3}(\bm{p}_2 - \bm{l}) \mbox{\hspace{0.2cm}}
\},
\eeqa
\end{widetext}
where $\B \Omega | \Omega \K = 1$ is used.

By assigning any of $\bm{i}$,$\bm{j}$,$\bm{k}$, and $\bm{l}$ in Eq.(\ref{AmpElement3}) to 
any of $\bm{s}$,$\bm{t}$,$\bm{u}$,$\bm{v}$ in Eq.(\ref{eq_TwoBodySmatrix}),
the two-body transition amplitude can be expressed as
\begin{widetext}
\beqa\label{eq_Amplitude}
\B \Omega^{'} | S^{(2)}_{2\rightarrow2} | \Omega \K =
\left(-\frac{1}{4}\frac{g}{M}\right)^2 \frac{i}{2}(2\pi)^4
\delta^{(4)}(p_1 + p_2 - p_3 - p_4) \sqrt{N_{p_1}} \sqrt{N_{p_2}} \sqrt{N_{p_4}} 
\times \mbox{\hspace{5cm}}\\ \nnb
\left(
\frac{8 (P_{1}P_{2})(\hat{P}_{3}\hat{P}_{4})}{m^2 - (p_1 + p_2)^2} +
\frac{4 (P_{2}\hat{P}_{4})(P_{1}\hat{P}_{3})}{m^2 - (p_2 - p_4)^2} +
\frac{4 (P_{2}\hat{P}_{3})(P_{1}\hat{P}_{4})}{m^2 - (p_2 - p_3)^2} +
\frac{4 (P_{1}\hat{P}_{4})(P_{2}\hat{P}_{3})}{m^2 - (p_1 - p_4)^2} +
\frac{4 (P_{1}\hat{P}_{3})(P_{2}\hat{P}_{4})}{m^2 - (p_1 - p_3)^2}
\right),
\eeqa
\end{widetext}
where subscripts have been omitted in the momentum-polarization tensors 
such as $(S\hat{T}) \equiv S_{\mu\nu}\hat{T}^{\mu\nu}$.

From the experimental point of view, it is also useful to consider the case
$p_1 + p_1 \rightarrow p_3 + p_4$, where the initial state photons are from a degenerate state,
because the number of incident beams can be reduced from two to one in an experimental setup.
For the degenerate case, we define the initial and final states, respectively, as follows:
\beqa
|\Omega\K &\equiv& |N_{\bm{p}_1}\K\K |N_{\bm{p}_4}\K\K |0\K, \mbox{and}\\ \nnb
\B \Omega^{'}| &\equiv& \B\B N_{\bm{p}_1}| \B\B N_{\bm{p}_4}| \B 1_{\bm{p}_3}| 
= \B \Omega | a_{\bm{p}_3}.
\eeqa
For the evaluation of the two-body transition amplitude 
$\B \Omega^{'} | S^{(2)}_{2\rightarrow2} | \Omega \K$
containing $a^{\dagger}_{\bm{i}} a^{\dagger}_{\bm{j}} a_{\bm{k}} a_{\bm{l}}$,
we again separately evaluate the contractions with coherent bra- and ket-states, respectively, as follows:
\beqa
\B \Omega^{'} | 
a^{\dagger}_{\bm{i}} a^{\dagger}_{\bm{j}}
= \sqrt{N_{p_1}}\hat{\delta}^{3}(\bm{p}_1 - \bm{i}) \times \mbox{\hspace{3.0cm}} \\ \nnb
\{ 
\B \Omega^{'} | \sqrt{N_{p_1}}\hat{\delta}^{3}(\bm{p}_1 - \bm{j}) + 
\B \Omega^{'} | \sqrt{N_{p_4}}\hat{\delta}^{3}(\bm{p}_4 - \bm{j}) + \\ \nnb
\B \Omega     | 1             \hat{\delta}^{3}(\bm{p}_3 - \bm{j})
\} + \\ \nnb
\sqrt{N_{p_4}}\hat{\delta}^{3}(\bm{p}_4 - \bm{i}) \times \mbox{\hspace{3.0cm}} \\ \nnb
\{ 
\B \Omega^{'} | \sqrt{N_{p_1}}\hat{\delta}^{3}(\bm{p}_1 - \bm{j}) +
\B \Omega^{'} | \sqrt{N_{p_4}}\hat{\delta}^{3}(\bm{p}_4 - \bm{j}) + \\ \nnb
\B \Omega     | 1             \hat{\delta}^{3}(\bm{p}_3 - \bm{j})
\} + \\ \nnb
1             \hat{\delta}^{3}(\bm{p}_3 - \bm{i}) \times \mbox{\hspace{3.5cm}} \\ \nnb
\{ 
\B \Omega | \sqrt{N_{p_1}}\hat{\delta}^{3}(\bm{p}_1 - \bm{j}) +
\B \Omega | \sqrt{N_{p_4}}\hat{\delta}^{3}(\bm{p}_4 - \bm{j}) + \\ \nnb
\B \Omega | a^{\dagger}_{\bm{j}}
\}, \mbox{\hspace{0.2cm}}
\eeqa
where the last term vanishes because $\B 0| a^{\dagger}_{\bm{p}} = 0$, and
\beqa
a_{\bm{k}} a_{\bm{l}} | \Omega \K = 
\sqrt{N_{p_1}}\hat{\delta}^{3}(\bm{p}_1 - \bm{l}) \times \mbox{\hspace{3.5cm}} \\ \nnb
\{ 
\sqrt{N_{p_1}}\hat{\delta}^{3}(\bm{p}_1 - \bm{k})|\Omega\K + 
\sqrt{N_{p_4}}\hat{\delta}^{3}(\bm{p}_4 - \bm{k})|\Omega\K
\} + \\ \nnb
\sqrt{N_{p_4}}\hat{\delta}^{3}(\bm{p}_4 - \bm{l}) \times \mbox{\hspace{3.5cm}} \\ \nnb
\{ 
\sqrt{N_{p_1}}\hat{\delta}^{3}(\bm{p}_1 - \bm{k})|\Omega\K + 
\sqrt{N_{p_4}}\hat{\delta}^{3}(\bm{p}_4 - \bm{k})|\Omega\K
\} \mbox{\hspace{0.3cm}}.
\eeqa

The consistent transition amplitude that satisfies the combination of the initial and final state
momenta in the degenerate case is expressed as
\beqa\label{AmpElement2}
\B \Omega^{'} | 
a^{\dagger}_{\bm{i}} a^{\dagger}_{\bm{j}} a_{\bm{k}}a_{\bm{l}}
| \Omega \K = 
\{ \sqrt{N_{p_4}}\hat{\delta}^{3}(\bm{p}_4 - \bm{i})1\hat{\delta}^{3}(\bm{p}_3 - \bm{j}) +
\mbox{\hspace{1.0cm}} \\ \nnb
1\hat{\delta}^{3}(\bm{p}_3 - \bm{i}) \sqrt{N_{p_4}}\hat{\delta}^{3}(\bm{p}_4 - \bm{j}) \} 
\times \mbox{\hspace{1.0cm}} \\ \nnb
\sqrt{N_{p_1}}\hat{\delta}^{3}(\bm{p}_1 - \bm{l})\sqrt{N_{p_1}}\hat{\delta}^{3}(\bm{p}_1 - \bm{k})
\B \Omega | \Omega \K \mbox{\hspace{0.0cm}} \\ \nnb 
= \sqrt{N_{p_1}} \sqrt{N_{p_1}} \sqrt{N_{p_4}} \times \mbox{\hspace{3.2cm}} \\ \nnb
\{
\hat{\delta}^{3}(\bm{p}_4 - \bm{i})\hat{\delta}^{3}(\bm{p}_3 - \bm{j})
\hat{\delta}^{3}(\bm{p}_1 - \bm{k})\hat{\delta}^{3}(\bm{p}_1 - \bm{l}) + \\ \nnb
\hat{\delta}^{3}(\bm{p}_3 - \bm{i})\hat{\delta}^{3}(\bm{p}_4 - \bm{j})
\hat{\delta}^{3}(\bm{p}_1 - \bm{k})\hat{\delta}^{3}(\bm{p}_1 - \bm{l})
\},
\eeqa
where $\B \Omega | \Omega \K = 1$ is substituted.

Again assigning any of $\bm{i}$,$\bm{j}$,$\bm{k}$, and $\bm{l}$ in Eq.(\ref{AmpElement2}) to 
any of $\bm{s}$,$\bm{t}$,$\bm{u}$, and $\bm{v}$ in Eq.(\ref{eq_TwoBodySmatrix}),
the two-body transition amplitude is expressed as
\begin{widetext}
\beqa\label{eq_AmplitudeDegenerate}
\B \Omega^{'} | S^{(2)}_{2\rightarrow2} | \Omega \K =
\left(-\frac{1}{4}\frac{g}{M}\right)^2
\frac{i}{2}(2\pi)^4 \delta^{(4)}(p_1 + p_1 - p_3 - p_4)
\sqrt{N_{p_1}} \sqrt{N_{p_1}} \sqrt{N_{p_4}} 
\times \mbox{\hspace{5cm}}\\ \nnb
\left(
\frac{4 (P_{1}P_{1})(\hat{P}_{3}\hat{P}_{4})}{m^2 - (p_1 + p_1)^2} +
\frac{4 (P_{1}\hat{P}_{4})(P_{1}\hat{P}_{3})}{m^2 - (p_1 - p_4)^2} +
\frac{4 (P_{1}\hat{P}_{3})(P_{1}\hat{P}_{4})}{m^2 - (p_1 - p_3)^2}
\right).
\eeqa
\end{widetext}
We take special note that
the degenerate case may also be interpreted as a special case of the non-degenerate case
by reducing the average number of $p_1$ photons from $N_{p_1}$ to $N_{p_1}/2$
due to the equal split into two identical beams and equating $1$ and $2$ in the subscripts
in Eq.(\ref{eq_Amplitude}).

\subsection{Kinematics in asymmetric-incident and non-coaxial geometry in QPS}
\begin{figure}[!hbt]
\centering
\includegraphics[width=0.40\textwidth]{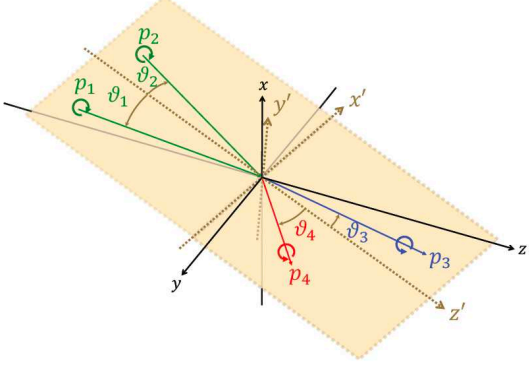}
\caption{Kinematical parameters in asymmetric-incident and non-coaxial geometry.
Primed coordinates $(x^{'}, y^{'}, z^{'})$, where the total transverse momentum of an arbitrary selected
pair of incident photons within the creation beam vanishes, are configurable with respect to 
the fixed laboratory coordinate $(x,y,z)$ to which the focused beam fields are physically mapped.
The kinematical parameters used to derive the scattering amplitude are all based
on these primed coordinates.
}
\label{Fig4}
\end{figure}
As illustrated in Fig.\ref{Fig4}, we extend the scattering formulation to the most general scattering
geometry, which is asymmetric-incident and non-coaxial scattering in QPS.
For a selected pair of incident waves, $p_1$ and $p_2$, from a coherent creation beam, 
we can always define an axis $z^{'}$ around which the total transverse momentum, $p_T$, 
of the two incident waves becomes zero and an axis $x^{'}$ normal to the $z^{'}$-axis
on the reaction plane that includes the two wave vectors, as shown in Fig.\ref{Fig4}.
On this $x^{'}-z^{'}$ plane, referred to as the {\it zero-$p_T$ coordinate}, 
the scattering amplitude is greatly simplified
because the emission angles of final state waves $p_3$ and $p_4$ become 
axially symmetric around the $z^{'}$-axis.
The labels for energies are common to both the zero-$p_T$ coordinate and
the laboratory coordinate defined in terms of the $x$,$y$, and $z$-axes,
while angular or momentum labels are valid only for the zero-$p_T$ coordinate.
In the following subsections, for simplicity,
we use kinematical parameters defined 
in the zero-$p_T$ coordinate, even when the prime symbol is not written, except where laboratory coordinates are explicitly specified.

With the energies of four photons $\om_i$ and scattering angles 
$\vth_i$ for initial $i=1,2$ and final $i=3,4$ states in the zero-$p_T$ coordinate,
four-momenta are defined as follows:
\begin{equation}
\label{momentum}
\begin{alignedat}{10}
p_{1} & = (& \om_{1},\ & & \om_{1}\sin\vth_{1},\ & & 0,\ & & \om_{1}\cos\vth_{1})&, \\
p_{2} & = (& \om_{2},\ & & -\om_{2}\sin\vth_{2},\ & & 0,\ & & \om_{2}\cos\vth_{2})&, \\
p_{3} & = (& \om_{3},\ & & \om_{3}\sin\vth_{3},\ & & 0,\ & & \om_{3}\cos\vth_{3})&,\\
p_{4} & = (& \om_{4},\ & & -\om_{4}\sin\vth_{4},\ & & 0,\ & & \om_{4}\cos\vth_{4}).&
\end{alignedat}
\end{equation}
For later convenience, a bisecting angle $\vth_b$ is introduced, with the meaning
\beqa\label{eq_thb}
\vth_{b} \equiv \frac{\vth_1+\vth_2}{2}.
\eeqa

The energy-momentum conservation equalities are
\begin{align}\label{eq_EnergyMomentum}
\om_1+\om_2 & = \om_3+\om_4\\ \nnb
\om_1\cos\vth_1+\om_2\cos\vth_2 & = \om_3\cos\vth_3+\om_4\cos\vth_4 \equiv \om_z\\ \nnb
\om_1\sin\vth_1-\om_2\sin\vth_2 & = \om_3\sin\vth_3-\om_4\sin\vth_4 \equiv \om_x
\end{align}
The corresponding center-of-mass energy, $E_{cms}$, is then expressed as
\begin{align}\label{eq_Ecms}
E_{cms} = \sqrt{2\om_1\om_2\{1-\cos(\vth_1+\vth_2)\}}= 2\sqrt{\om_1\om_2}\sin\vth_b{}.
\end{align}

We then define the linear polarization vectors as
\beqa
\bm{e}_i^{(1)} &=& (0, 1, 0), \\ \nnb
\bm{e}_1^{(2)} &=& (-\cos\vth_1, 0, \sin\vth_1), \\ \nnb
\bm{e}_2^{(2)} &=& (-\cos\vth_2, 0, -\sin\vth_2), \\ \nnb
\bm{e}_3^{(2)} &=& (-\cos\vth_3, 0, \sin\vth_3),  \\ \nnb
\bm{e}_4^{(2)} &=& (-\cos\vth_4, 0, -\sin\vth_4).
\eeqa
With the linear polarization vectors, we also can define circular polarization states:
\beqa
\bm{e}^R_i = \frac{1}{\sqrt{2}}(\bm{e}_i^{(1)} + i\bm{e}_i^{(2)}), \\ \nnb
\bm{e}^L_i = \frac{1}{\sqrt{2}}(\bm{e}_i^{(1)} - i\bm{e}_i^{(2)}).
\eeqa

Given these definitions, we can evaluate the momentum-polarization tensors 
included in Eq.(\ref{eq_Amplitude}) for the circular polarization case as follows
\beqa\label{eq_cir}
(P_iP_j) = (\hat{P}_i\hat{P}_j) &=& 2\om_i\om_j(1-\cos(\vartheta_i+\vartheta_j)) \\ \nnb
(P_i\hat{P}_j) = (\hat{P}_iP_j) &=& 0.
\eeqa

\subsection{Lorentz-invariant scattering amplitude including a resonance state}
Here we are particularly interested in formulating a Lorentz-invariant scattering
amplitude for circular polarization states,
because the states describe naturally interpretable angular momenta of photons 
with respect to any directions of the photon momenta. 
We denote a sequence of four-photon circular polarization states
as a subscript $S\equiv abcd$ with $a,b,c,d =$ $R$ (right-handed) or $L$ (left-handed).
From the following definition for the transition amplitude
\beqa
\B \Omega^{'} | S^{(2)}_{2  \rightarrow 2}|\Omega \K =
\sqrt{N_{p_1}}\sqrt{N_{p_2}}\sqrt{N_{p_4}} \times \\ \nnb
i(2\pi)^4 \delta^{(4)}(p_1 + p_2 - p_3 - p_4) {\cal M}_S,
\eeqa
the Lorentz-invariant scattering amplitude ${\cal M}_S$ can be expressed as
\beqa
{\cal M}_S = \frac{1}{4} \left(\frac{g}{M}\right)^2 
\frac{(P_{1}P_{2})(\hat{P}_{3}\hat{P}_{4})}{m^2-(p_1+p_2)^2}
\eeqa
from Eq.(\ref{eq_Amplitude}) by taking Eq.(\ref{eq_cir}) into account,
where we exclude the beam-relevant factor in the definition of ${\cal M}_S$ so as to
decouple the dynamics from the experimental factor caused by the coherent beam intensity.
With Eq.(\ref{eq_cir}), the vertex factors in the numerator of ${\cal M}_S$ are expressed as
\beqa
(P_1P_2) &=& 2\om_1\om_2(1-\cos(\vartheta_1+\vartheta_2)) \\ \nnb
(\hat{P}_3\hat{P}_4) &=& 2\om_3\om_4(1-\cos(\vartheta_3+\vartheta_4)).
\eeqa
Since energy-momentum conservation requires $(p_1 + p_2)^2 = (p_3 + p_4)^2$,
we can describe the amplitude applicable to both $S=LLRR$ and $RRLL$
in terms of only the initial state variables, as follows:
\beqa\label{eq_Ms}
{\cal M}_S &=& \left(\frac{g}{M}\right)^2 
\frac{(\om_1\om_2(\cos(\vth_1+\vth_2)-1))^2}{m^2-2\om_1\om_2(1-\cos(\vth_1+\vth_2))} \quad \\ \nnb
&=& \left(\frac{g}{M}\right)^2 
\frac{(2\om_1\om_2\sin^2\vth_b)^2}{m^2-4\om_1\om_2\sin^{2}\vth_b},
\eeqa
where Eq.(\ref{eq_thb}) is substituted for the second relation.
To implement energy fluctuations in the initial state of two photons chosen
from a solo coherent beam around its central energy $\om_c$, we introduce two independent
parameters $s_1$ and $s_2$, as follows:
\beq\label{eq_s1s2}
\om_1 = s_1 \om_{\mrm{c}},\quad
\om_2 = s_2 \om_{\mrm{c}}.
\eeq
We then define a resonance energy $\om_r$ satisfying $E_{cms} = m$ as
\beq\label{eq_omr}
\om^2_r \equiv \frac{m^2}{4s_1s_2\sin^2\vth_b}.
\eeq
Because the exchanged scalar field is (in principle) an unstable particle, we introduce a decay rate
$\Gamma$, which is defined as~\cite{PTP-DE}
\beq
\Gamma = \frac{1}{16\pi}\left( \frac{g}{M} \right)^2 m^3.
\eeq
This causes a change in the mass square as $m^2 \rightarrow (m - i\Gamma/2)^2 \approx m^2 - i m\Gamma$.
Therefore, the denominator $\mcal{D}$ in Eq.(\ref{eq_Ms}) is expressed as
\begin{align}
\label{eq_D}
\mcal{D}
& \approx -4s_{1}s_{2}\om_{\mrm{c}}^{2}\sin^{2}\vth_{b}+m^{2}-i\mathit{\Gamma}m \nnb \\
& = -4s_{1}s_{2}\om_{\mrm{c}}^{2}\sin^{2}\vth_{b}+4s_{1}s_{2}\om_{r}^{2}\sin^{2}\vth_{b}-i\mathit{\Gamma}m \nnb \\
& = -4s_{1}s_{2}\sin^{2}\vth_{b} \left( \om_{\mrm{c}}^{2}-\om_{r}^{2}+i\frac{\mathit{\Gamma}m}
			{4s_{1}s_{2}\sin^{2}\vth_{b}} \right) \nnb \\
& \equiv -4s_{1}s_{2}\sin^{2}\vth_{b}(\chi+ia),
\end{align}
where 
\beq\label{eq_chi}
\chi \equiv \om^2_c - \om^2_r = \left( 1-\frac{m^2}{4\om_1\om_2\sin^2_{\vth_b}} \right)\om^2_c
\eeq
describes the degrees of deviation of $E_{cms}$ as determined 
by a pair of incident photons from the resonance energy derived from the central energy $\om_c$, 
and $a$ is defined as
\beq\label{eq_a}
a = \frac{\mathit{\Gamma}m\om^2_c}{4\om_1\om_2 \sin^2 \vth_b} 
  = \frac{1}{16\pi} \left(\frac{g}{M}\right)^2 \om^2_r m^2.
\eeq
The numerator $\mcal{N}$ in Eq.(\ref{eq_Ms}) is considered to be
\beqa
\label{Numerator}
\mcal{N} &=& \left(\frac{g}{M}\right)^2 (2\om_1\om_2\sin^{2}\vth_{b})^{2} \\ \nnb
&=& 16\pi as_{1}s_{2}\sin^{2}\vth_{b} \left(\frac{\om_c}{\om_r}\right)^4
\sim 16\pi as_{1}s_{2}\sin^{2}\vth_{b},
\eeqa
where Eq.(\ref{eq_omr}) and (\ref{eq_a}) are used for the second expression
and the condition $\om_c \sim \om_r$ is required for the last step 
because the resonance condition $E_{cms}=m$ can be satisfied dominantly
with a proper choice of $\vth_b$ for a given $m$ 
without changing the central beam energy $\om_c$ itself.
Finally, the following Breit--Wigner distribution is obtained:
\beq
\label{Ms_squared}
\mcal{M}_{s} = \frac{\mcal{N}}{\mcal{D}} = 4\pi\frac{a}{\chi+ia},\quad \mbox{and} \quad
|\mcal{M}_{s}|^{2} = (4\pi)^{2}\frac{a^{2}}{\chi^{2}+a^{2}}.
\eeq

Since we expect that $E_{cms}$ is in principle uncertain due to unavoidable
energy and momentum uncertainties of a selected pair of two photon wave vectors in QPS,
we need to average the resonance effect over a range of $\chi$.
In order to show the essence of inclusion of an resonance state within a range from 
$\chi_{-}$ to $\chi_{+}$, we demonstrate the simplest averaging as follows.
We define $\chi_{\pm}$ in units of $a$ as $\chi_{\pm} = \pm \eta a$ with $\eta \gg 1$.
The averaging process is expressed as
\beqa
\overline{|{\cal M}_s|^2} &=& \frac{1}{\chi_+ - \chi_-}\Int{\chi_-}{\chi_+} 
|{\cal}M_s|^2 d\chi \\ \nnb
&=& \frac{(4\pi)^2}{2\eta a} 2a \tan^{-1}(\eta) = (4\pi)^2\eta^{-1}\tan^{-1}(\eta) \\ \nnb
&\approx& (4\pi)^2\eta^{-1}\frac{\pi}{2} = 8\pi^2\frac{a}{|\chi_{\pm}|},
\eeqa
with the approximation due to $\eta \gg 1$.
Compared to non s-channel cases where $|{\cal M}_s|^2 \propto a^2$,
capturing a resonance within the $E_{cms}$ uncertainty has 
a gain of $a^{-1} \propto M^2$ as shown above.
If the energy scale $M$ corresponds to the Planckian scale $M_p$, this gain factor is huge
even if we cannot directly capture the top of the Breit-Wigner distribution 
where $|{\cal M}_s|^2 \propto (4\pi)^2$ with $\chi \rightarrow 0$.
This is the prominent feature of s-channel scattering including a resonance in QPS.
In the following subsections, we will introduce more realistic probability distribution functions
for $E_{cms}$ based on the physical nature of propagating electromagnetic fields 
in order to implement the averaging process.

\subsection{Evaluation of signal yield in stimulated resonant scattering}
Let us first consider the number of scattering events in $p_1 + p_2 \rightarrow p_3 + p_4$
with two colliding photon beams having normalized densities $\rho_1$ and $\rho_2$ with 
average number of photons $N_1$ and $N_2$, respectively. 
This is referred to as the {\it spontaneous} yield to get the signal $p_3$ in the final state.
With the Lorentz-invariant phase space factor $dL_{ips}$
\beq\label{eqLIPS}
dL_{ips} = (2\pi)^4 \delta(p_3+p_4-p_1-p_2)
\frac{d^3p_3}{2\omega_3(2\pi)^3}\frac{d^3p_4}{2\omega_4(2\pi)^3},
\eeq
the spontaneous signal yield ${\cal Y}$ can be factorized according to the concept of
time-integrated luminosity ${\cal L}$ times {\it cross section} $\sigma$,
as follows:
\beqa\label{eqYsigma}
{\mathcal Y} = N_1 N_2
\left( \int dt d\bm{r} \rho_1(\bm{r},t) \rho_2(\bm{r},t) K(p_1,p_2) \right) 
\times \mbox{\hspace{1.5cm}} \\ \nnb
\left(
\frac{c}{2\om_1 2\om_2 K(p_1,p_2)}
|{\mathcal M}_s(p_1,p_2)|^2 dL_{ips}
\right)
\mbox{\hspace{1.1cm}} \\ \nnb
\equiv
{\mathcal L(p_1,p_2)}\left[s \cdot L^3 \cdot L^{-3} \cdot L^{-3} \cdot L/s)\right]
\sigma(p_1,p_2)\left[L^2\right], \mbox{\hspace{0.5cm}}
\eeqa
where $K$ corresponds to the relative velocity of the incoming particle beams
between two incident photons with velocity vectors $\bm{v_1}$ and $\bm{v_2}$,
based on M{\o}ller's Lorentz-invariant factor~\cite{Moeller}.
The relative velocity $K$ is defined as~\cite{K-factor}
\beq\label{eqK}
K(p_1,p_2) \equiv
\sqrt{(\bm{v_1}-\bm{v_2})^2 -
\frac{(\bm{v_1} \times \bm{v_2})^2}{c^2}}
\eeq
with $c$ the velocity of light. The notation $[\quad]$ indicates units with length $L$ and time $s$.
 
The concept of the {\it cross section} is convenient for fixed $p_1$ and $p_2$ beams.
However, in order to implement fluctuations on the velocity vectors, which are
represented by the integral on the probability density of cms-energy
$W(Q)$ as a function of the combinations of energies and angles---in laboratory coordinates,
denoted as 
\beq\label{eq_QdQ}
Q \equiv \{\omega_{\alpha}, \Theta_{\alpha}, \Phi_{\alpha}\} \quad \mbox{and} \quad
dQ \equiv \Pi_{\alpha} d\omega_{\alpha} d\Theta_{\alpha} d\Phi_{\alpha} 
\eeq
for the incident beams $\alpha=1, 2$-- the
{\it volume-wise interaction rate} $\overline{\Sigma}$ defined below~\cite{BJ}
is more straightforward than the {\it cross section} $\sigma$
\beqa\label{eq_Y}
{\mathcal Y} = N_1 N_2
\left(
\int dt d\bm{r} \rho_1(\bm{r},t) \rho_2(\bm{r},t)
\right)
\times \mbox{\hspace{2cm}} \\ \nnb
\left(
\int dQ W(Q)
\frac{c}{2\om_1 2\om_2} |{\mathcal M}_s(Q^{'})|^2 dL^{'}_{ips}
\right)
\mbox{\hspace{0.2cm}} \nnb\\
\equiv N_1 N_2 {\mathcal D}\left[s/L^3\right] \overline{\Sigma}\left[L^3/s\right],
\mbox{\hspace{3.3cm}}
\eeqa
because the intermediate $K$-factor is canceled in advance of averaging
over $W(Q)$, where $Q^{'} \equiv \{\omega_{\alpha}, \vth_{\alpha}, \phi_{\alpha}\}$
are kinematical parameters used for the zero-$p_T$ coordinate constructed from
a pair of two incident waves.
The conversions from $Q$ to $Q^{'}$ are possible through rotation functions 
$\vth_{\alpha} \equiv {\cal R}_{\vth_{\alpha}}(Q)$
and
$\phi_{\alpha} \equiv {\cal R}_{\phi_{\alpha}}(Q)$.

We then extend the {\it spontaneous} yield to the {\it induced} yield, ${\cal Y}_I$,
by adding one more beam with the central four-momentum $p_4$ having normalized density $\rho_4$
with the extended set of parameters: 
\beq\label{eq_QI}
Q_I \equiv \{Q, \omega_4, \Theta_4, \Phi_4\} \quad \mbox{and} \quad
dQ_I \equiv dQd\omega_4 d\Theta_4 d\Phi_4.
\eeq
The induced yield is then expressed as
\beqa\label{eq_YI}
{\mathcal Y}_I = N_1 N_2 N_4
\left(
\int dt d\bm{r} \rho_1(\bm{r},t) \rho_2(\bm{r},t) \rho_4(\bm{r},t) V_4
\right) \times \mbox{\hspace{0.7cm}} \\ \nnb
\left(
\int dQ_I W(Q_I)
\frac{c}{2\om_1 2\om_2} |{\mathcal M}_s(Q^{'})|^2 dL^{'I}_{ips} \mbox{\hspace{0.1cm}}
\right)
\\ \nnb
\equiv  N_1 N_2 N_4 {\mathcal D}_I\left[s/L^3\right] \overline{\Sigma}_I\left[L^3/s\right],
\mbox{\hspace{3.1cm}}
\eeqa
where $\rho_4(\bm{r},t) V_4$ with the volume of the $p_4$ beam, $V_4$, corresponds to the probability 
density that describes a spacetime overlap of the $p_1$ and $p_2$ beams with the inducing beam $p_4$; 
$dL^{'I}_{ips}$ indicates the inducible phase space in which the 
solid angles of $p_3$ must be consistent so that the balancing solid angles of $p_4$
determined via energy-momentum conservation can be found within the distribution
of the given inducing beam (in laboratory coordinates) 
after conversion from $p_4$ in the zero-$p_T$ coordinate system to 
the corresponding laboratory coordinate. $W(Q_I)$ is explicitly introduced as
\beq\label{eq_WQI}
W(Q_I) \equiv \Pi_\beta G_E(\omega_{\beta}) G_p(\Theta_{\beta},\Phi_{\beta})
\eeq
with Gaussian distributions $G$ for
\beq\label{eq_dQI}
dQ_I \equiv \Pi_{\beta} d\omega_{\beta} d\Theta_{\beta} d\Phi_{\beta}
\eeq
over $\beta = 1, 2, 4$.
The Gaussian distributions $G_E$ in the energy space and $G_p$ in the momentum space 
(equivalently, the polar angles in the case of photons) are introduced according to the properties of 
a focused coherent electromagnetic field with an axial symmetric nature for an 
azimuthal angle $\Phi$ around the optical axis of a focused beam, as we discuss soon.

We then specifically consider a search for signal photons $p_3$ for the degenerate case 
in the generic QPS including asymmetric collisions: $p_c + p_c \rightarrow p_3 + p_4$
where $p_1$ and $p_2$ are stochastically obtained from a single focused coherent beam with 
central photon energy $\om_c$ under the co-propagating focused coherent beam 
with central four-momentum $p_4$ for the purpose of induction.
Based on the transition amplitude in Eq.(\ref{eq_AmplitudeDegenerate})
and the yield expression in Eq.(\ref{eq_YI}), the induced signal yield in the degenerate case,
where we combine a coherent creation beam and a coherent inducing beam
with the average numbers of photons $N_c$ and $N_i$, respectively, 
and focus them into the common optical axis (in laboratory coordinates) 
is expressed as
\beqa\label{eq_Yci}
{\mathcal Y}_{c+i} = (N_c/2) (N_c/2) N_i \times \mbox{\hspace{4cm}} \\ \nnb
\left(
\int dt d\bm{r} \rho_c(\bm{r},t) \rho_c(\bm{r},t) \rho_i(\bm{r},t) V_i
\right)
\times \mbox{\hspace{0.8cm}} \\ \nnb
\left(
\int dQ_I W(Q_I)
\frac{c}{2\om_1 2\om_2} |{\mathcal M}_s(Q^{'})|^2 dL^{'I}_{ips} \mbox{\hspace{0.1cm}}
\right)
\\ \nnb
\equiv  \frac{1}{4} N^2_c N_i {\mathcal D}_I\left[s/L^3\right] \overline{\Sigma}_I\left[L^3/s\right],
\mbox{\hspace{2.3cm}}
\eeqa
where the factor $1/4$ appears for the reason explained in the paragraph
just below Eq.(\ref{eq_AmplitudeDegenerate}) 
and ${\cal M}_S$ is based on the non-degenerate case resulting in Eq.(\ref{eq_Ms}). 
In the following, we provide detailed formulas for ${\cal D}_I$
and $\overline{\Sigma}_I$ in Eq.(\ref{eq_Yci}).

\subsubsection{\bf Properties of a Gaussian beam in vacuum}\label{ss_Gaussian}
The solution for propagation of an electromagnetic field in vacuum is
known as the basic Gaussian mode~\cite{Yariv}.
In the Gaussian mode, the electric field propagating along the $z$-direction with wave number $k$
in (laboratory) spatial coordinates $(x, y, z)$ is expressed as
\beqa\label{eq_EGauss}
\bm{E}(x, y, z)
= \bm{E}_{0}\frac{w_{0}}{w(z)} \times \mbox{\hspace{4cm}} \\ \nnb
\exp{\left( -i(kz-\eta(z))-r^{2}
\left( \frac{1}{w^{2}(z)}+\frac{ik}{2R(z)} \right) \right)},
\eeqa
where the individual factors are summarized as follows.
\beq
\begin{split}
r & = \sqrt{x^{2}+y^{2}}\\
w(z) & = w_{0}\sqrt{1+\frac{z^{2}}{z_{R}^{2}}}\\
\eta(z) & = \tan^{-1}\left( \frac{z}{z_{R}} \right)\\
R(z) & = z\left(1+\frac{z_{R}^{2}}{z^{2}} \right)
\end{split}
\eeq
In this, the beam waist $w_0$ and Rayleigh length $z_R$ are 
\beq
w_0 = \frac{\lambda}{\pi\vth_0}, \quad z_R = \frac{\pi w^2_0}{\lambda}
\eeq
for a given wavelength $\lambda$.
When a single electromagnetic field is focused with focal length $f$ and beam diameter $d$,
the beam waist is related to the incident angle $\Theta_0$ by
\beq\label{eq_Theta0}
\Theta_{0} = \tan^{-1}\left( \frac{d}{2f} \right).
\eeq
At the focal point $z=0$, the spatial distribution of the electric field is expressed as
\beq
\bm{E}(x, y, z=0) = \bm{E}_{0}\exp{\left( -\frac{x^{2}+y^{2}}{w_{0}^{2}} \right)}.
\eeq
The corresponding wave number distribution is obtained 
by Fourier transformation of the electric field, yielding
\beqa\label{eq_kT}
\hat{\bm{E}}(k_{x}, k_{y}, 0)
= \frac{1}{4\pi^{2}}\Int{-\infty}{\infty}\bm{E}_{0}\exp{\left( -\frac{x^{2}+y^{2}}{w_{0}^{2}} \right)}
\times \mbox{\hspace{1.5cm}} \\ \nnb
\exp(-i(k_{x}x+k_{y}y))dxdy
= \frac{w_{0}^{2}\bm{E}_{0}}{4\pi}\exp{\left( -\frac{w_{0}^{2}}{4}(k_{x}^{2}+k_{y}^{2}) \right)}.
\eeqa
The uncertainty on incident angles of wave vectors within the electric field
with respect to the $z$-axis 
can be related to $k_T = \sqrt{k^2_x + k^2_y}$ via the variance
\beq\label{eq_sigmakT}
\sigma^2_{k_T} = \frac{2}{w^2_0}
\eeq
in the Gaussian form.

For incident angles $\Theta$ defined with the transverse momenta $k_T$ 
in Eq.(\ref{eq_sigmakT}) and the incident energies $\om$ with $\hbar=1$,
\beq
\Theta = \sin^{-1}\left(\frac{k_T}{\om}\right),
\eeq
and the error propagation on the incident angles is given by
\begin{align}\label{eq_AngleSpread}
\sigma_{\Theta}
& = \sqrt{ \left( \pdiff{\Theta}{k_T} \right)^2\sigma_{k_T}^2
+\left( \pdiff{\Theta}{\om} \right)^2\sigma_{\om}^2 }\\ \nnb
& = \sqrt{ \frac{1}{\B \om \K^2-\B k_T \K^2}
\left( \sigma_{k_T}^2+\left( \frac{\B k_T \K}{\B \om \K} \right)^2
\sigma^2_{\om} \right) } \\ \nnb
& = \frac{1}{\om_c}\sigma_{k_T} = \frac{1}{\sqrt{2}}\Theta_0,
\end{align}
where $\B \om \K = \om_c$, $\B k_T \K = 0$, and 
Eq.(\ref{eq_sigmakT}) are substituted for the last line.

The average number of photons, $N$, in a pulsed electromagnetic field can be related to the
square of the electric field, $I$, by adding a Gaussian-shaped time distribution with
duration $\tau$, as follows:
\beq
I(t,x^{i}) = E^2_0 \frac{w_0^{2}}{w^2(ct)}
\exp\left( -2\frac{x^{2}+y^{2}}{w^2(ct)} \right)
\exp\left( -2\left( \frac{z-ct}{c\tau} \right)^{2} \right),
\eeq
where $E^2_0$ corresponds to $N$.
The volume for the normalization is then expressed as
\beq\label{eq_V}
V = \Int{-\infty}{\infty}dx^{i} \frac{I}{E^2_0}
= \left( \frac{\pi}{2} \right)^{\frac{3}{2}}w^2_0c\tau.
\eeq
Therefore, the normalized density profile per photon, $\rho \equiv I/(NV)$, is expressed as
\beq\label{eq_rho}
\rho(t,x^{i}) = 
\frac{(2/\pi)^{3/2}}{w^2(ct) c\tau}
\exp\left( -2\frac{x^{2}+y^{2}}{w^2(ct)} \right)
\exp\left( -2\left( \frac{z-ct}{c\tau} \right)^{2} \right).
\eeq

\subsubsection{\bf Integrated inducible volume-wise interaction rate, $\overline{\Sigma}_I$}
With the kinematical parameters defined in a zero-$p_T$ coordinate as illustrated in Fig.\ref{Fig4},
$Q^{'} \equiv \{\om_1,\om_2,\vth_1, \vth_2, \phi_1, \phi_2\}$, 
we first discuss the integrand of the spontaneous volume-wise interaction rate in Eq.(\ref{eq_Y})
in individual zero-$p_T$ coordinates
\beq
\Sigma^{'} \equiv \frac{c}{2\om_1 2\om_2} |{\mathcal M}_{s}(Q^{'})|^2 dL^{'}_{ips}.
\eeq
With $d^3 p_3 = \om^2_3 d\om_3 d\Omega^{'}_3$, the differential volume-wise interaction rate
per solid angle $d\Omega^{'}_3$ in a zero-$p_T$ coordinate is expressed as
\beqa\label{eq_dSigma}
\diff{\Sigma^{'}}{\mathit{\Omega}^{'}_{3}}
= \frac{c}{32\pi^{2}\om_1\om_2} \Int{0}{\infty} d\om_3\om_3 \times \mbox{\hspace{3cm}} \\ \nnb
\Int{-\infty}{\infty} d^{3}dp_4
\frac{|\mcal{M}_{s}(Q^{'})|^2}{2\om_4}\delta^{4}(p_3 + p_4 - p_1 - p_2).
\eeqa
We then insert the following identity
\beq\label{identity}
1 = \Int{0}{\infty}dp_{4}^{0} \delta(p_{4}^{0}-\om_{4})
= \Int{-\infty}{\infty}dp_{4}^{0} 2\om_{4}\delta(p_{4}^{2})\mathit{\Theta}(p_{4}^{0})
\eeq
to derive
\beqa
\int\frac{d^3p_4}{2\om_4}\delta^4(p_3 + p_4 - p_1 - p_2) \mbox{\hspace{3cm}} \\ \nnb
= \int d^4p_4\delta(p^2_4)\mathit{\Theta}(p^0_4)\delta^4(p_3 + p_4 - p_1 - p_2) \\ \nnb
=\delta((p_1 + p_2 - p_3)^2), \mbox{\hspace{3.25cm}}
\eeqa
where $\om_4 > 0$ is guaranteed.
In an asymmetric QPS, the following relation holds due to energy-momentum conservation:
\begin{align}
\label{p4_squared}
p_{4}^{2}
& = (p_{1}+p_{2}-p_{3})^{2} \nnb \\
& = 2(\om_{1}+\om_{2}-\om_{z}\cos\vth_{3})
\left( \om_{3}-\frac{2\om_{1}\om_{2}\sin^{2}\vth_{b}}{\om_{1}+\om_{2}-\om_{z}\cos\vth_{3}} \right).
\end{align}
Hence,
\begin{align}
\label{delta_p4_squared}
\delta(p_{4}^{2})
& = \frac{1}{2(\om_{1}+\om_{2}-\om_{z}\cos\vth_{3})} \delta(\om_{3}-\hat{\om}_{3}) \nnb \\
& = \frac{\hat{\om}_{3}}{4\om_{1}\om_{2}\sin^{2}\vth_{b}} \delta(\om_{3}-\hat{\om}_{3}),
\end{align}
where 
\beq
\label{hatom3_def}
\hat{\om}_{3}
\equiv \frac{2\om_{1}\om_{2}\sin^{2}\vth_{b}}{\om_{1}+\om_{2}-\om_{z}\cos\vth_{3}}.
\eeq

Therefore, we can write the expression
\begin{align}\label{diffCS_Om3}
\diff{\Sigma^{'}}{\mathit{\Omega}^{'}_{3}}
=\frac{c}{32\pi^{2}\om_{1}\om_{2}} \times \mbox{\hspace{5cm}} \\ \nnb
\Int{0}{\infty}d\om_{3}\om_{3}
\frac{\hat{\om}_{3}}{4\om_{1}\om_{2}\sin^{2}\vth_{b}}|\mcal{M}_{s}(Q^{'})|^{2}
\delta(\om_{3}-\hat{\om}_{3}) \\ \nnb
=\frac{c\hat{\om}_{3}^{2}|\mcal{M}_{s}(Q^{'})|^{2}}{2(8\pi\om_{1}\om_{2}\sin\vth_b)^2}.
\mbox{\hspace{4.1cm}}
\end{align}

Because the incident energies and momenta fluctuate for the single creation beam,
the differential volume-wise interaction rate must be averaged over possible values of
$\chi$ in Eq.(\ref{eq_chi}) according to the probability distribution functions
$W(Q) \equiv G_E(\om_1) G_E(\om_2) G_p(\Theta_1, \Phi_1) G_p(\Theta_2, \Phi_2)$
with the parameters in laboratory coordinates,
where
\beq
G_E(\om) \equiv 
\frac{1}{\sqrt{2\pi}\sigma_{\om}}\exp{\left(-\frac{(\om-\bar{\om})^2}{2\sigma^2_{\om}}\right)},
\eeq
with mean $\bar{\om}$
and
\beq
G_p(\Theta, \Phi) \equiv 
\frac{1}{2\pi\sigma^2_{\Theta}}
\exp \left( -\frac{\Theta^2}{2\sigma^2_{\Theta}} \right)
= \frac{1}{\pi\Theta^2_0} \exp \left( -\frac{\Theta^2}{\Theta^2_0} \right)
\eeq
by substituting $\sigma_{\Theta} = \Theta_0/\sqrt{2}$ from Eq.(\ref{eq_AngleSpread}) for the second.
We note that $G_p$ is normalized to the two-dimensional Gaussian distribution in $\Theta-\Phi$
angular space, where the $\Phi$-dependence is implicitly implemented 
via the axial symmetric feature of a focused beam even though the right-hand side includes
only the $\Theta$-dependence.
With the explicit notation $dQ \equiv d\om_1 d\om_2 d\Theta_1 d\Theta_2 d\Phi_1 d\Phi_2$,
the integrated differential volume-wise interaction rate in the zero-$p_T$ coordinate is then expressed as
\begin{align}\label{eq_AvedSdO3}
\diff{\overline{\Sigma}^{'}}{\Omega^{'}_3}
\equiv \int dQ W (Q) \frac{c|\mcal{M}_{s}({\cal R}(Q))|^{2}}{2(8\pi\om_{1}\om_{2}\sin\vth_{b})^{2}},
\end{align}
where ${\cal R}$ denotes rotation functions that convert
a $Q$ given in laboratory coordinates to $Q^{'}$ in the corresponding zero-$p_T$ coordinate system.

So far, we have discussed the spontaneous scattering process resulting in the two-photon final state
with $p_3$ and $p_4$. We now discuss the stimulated volume-wise interaction rate with
a coherent inducing field at the spacetime where the scattering takes place. 
We then need to revisit the commutation relation 
used in Eq.(\ref{eq_commutation}). To have the enhancement factor $\sqrt{N_{p_4}}$
appear through the second relation in Eq.(\ref{eq_enhancement}), both momentum and polarization states
of the spontaneous $p_4$-wave must be identical with those in the inducing coherent field.
As for the matching of polarization state, as long as we consider circular polarization states
(for instance, $S=LLRR$), the matching is satisfied for any directions of $p_4$-waves 
in the inducing beam with the $R$-state resulting in a $p_3$-wave with an $R$-state. On the other hand,
for the momentum state matching, we need to evaluate what fraction of the inducing beam 
can actually stimulate the scattering process; that is, the enhancement factor possible for 
the coherent state, because the focused short-pulse inducing beam has a spread in 
both momentum and energy spaces.
Phase-space matching can be implemented by introducing the symbol $dL^{'I}_{ips}$. 
This symbol indicates that we take into account the solid angles of signal photons, $p_3$, 
only when we can find balancing $p_4$ waves via energy-momentum conservation 
within the given focused inducing field. More explicitly, we define the following relation:
\beqa
\overline{\Sigma}_I \equiv 
\int G_E(\om_4) G_p(\Theta_4, \Phi_4) d\Omega_4  \times \\ \nnb
\diff{\Omega^{'}_4}{\Omega_{4}}
\diff{\Omega{'}_3}{\Omega{'}_4} 
\diff{\overline{\Sigma}^{'}}{\Omega{'}_{3}},
\eeqa
where 
\beqa
\diff{\Omega^{'}_{3}}{\Omega^{'}_{4}} =
\frac{d\phi_3\sin\vth_3}{d\phi_4\sin\vth_4}\diff{\vth_3}{\vth_4}
= \frac{\sin\vth_3}{\sin\vth_4}\diff{\vth_3}{\vth_4}
= -\left(\frac{\om_4}{\om_3}\right)^2.
\eeqa
This is based on the energy-momentum conservation in Eq.(\ref{eq_EnergyMomentum})
and $d\phi_3=d\phi_4$ in zero-$p_T$ coordinates.
Because an inducible solid angle of $p_4$ in a zero-$p_T$ coordinate must
match with a solid angle within the angular distribution of the inducing coherent field
mapped in laboratory coordinates, $d\Omega_4 = d\Omega^{'}_4$ must be satisfied.
Therefore, the inducible volume-wise interaction rate is eventually expressed as
\beqa\label{eq_Sibar}
\overline{\Sigma}_I
= -\int G_E(\om_4) G_p(\Theta_4, \Phi_4) d\Omega_4 
\left(\frac{\om_4}{\om_3}\right)^2 
\diff{\overline{\Sigma}^{'}}{\Omega^{'}_{3}} \\ \nnb
= 
\Int{0}{2\pi} d\Phi_4 \Int{\pi/2}{0} d\Theta_4 \sin\Theta_4 G_p(\Theta_4, \Phi_4)
\times \mbox{\hspace{0.25cm}} \\ \nnb
\int d\om_4 G_E(\om_4)\left(\frac{\om_4}{\om_1+\om_2-\om_4}\right)^2
\times \mbox{\hspace{0.8cm}} \\ \nnb
\int dQ
W (Q) \frac{c|\mcal{M}_{s}({\cal R}(Q))|^{2}}{2(8\pi\om_{1}\om_{2}\sin\vth_{b})^{2}},
\mbox{\hspace{1.4cm}}
\eeqa
where all integral variables are expressed by those defined 
in the laboratory coordinate, $Q_I$ as defined 
in Eq.(\ref{eq_QI}) that includes $Q$ in Eq.(\ref{eq_QdQ}).

\subsubsection{\bf Spacetime overlapping factor with an inducing beam, ${\mcal D}_I$}
For scattering in QPS, we introduce a common normalized density for the incident beams
as $\rho_c \equiv \rho_1 = \rho_2$ by assuming that $p_1$ and $p_2$ are stochastically
selected from the single creation beam and the inducing beam $\rho_i$ 
based on Eq.(\ref{eq_rho}) as follows:
\beqa
\rho_{\mrm{c}}(t,x^{i})
 = \left( \frac{2}{\pi} \right)^{\frac{3}{2}}\frac{1}{w_{\mrm{c}}^{2}(ct)c\tau_{\mrm{c}}}
\times \mbox{\hspace{3.4cm}} \\ \nnb
\exp\left( -2\frac{x^{2}+y^{2}}{w_{\mrm{c}}^{2}(ct)} \right)
\exp\left( -2\left( \frac{z-ct}{c\tau_{\mrm{c}}} \right)^{2} \right)
\eeqa
\beqa
\rho_{\mrm{i}}(t,x^{i})
 = \left( \frac{2}{\pi} \right)^{\frac{3}{2}}\frac{1}{w_{\mrm{i}}^{2}(ct)c\tau_{\mrm{i}}}
\times \mbox{\hspace{3.4cm}} \\ \nnb
\exp\left( -2\frac{x^{2}+y^{2}}{w_{\mrm{i}}^{2}(ct)} \right)
\exp\left( -2\left(\frac{z-ct}{c\tau_{\mrm{i}}} \right)^{2} \right),
\eeqa
where the origin is defined at $t=0$ and $x^i = 0$ for $i=1-3$, and the pulse durations $\tau_c$ and $\tau_i$ 
follow independent Gaussian distributions.
Because we discussed the angular spread only at the origin in Eq.(\ref{eq_kT}), 
where the maximum interaction rate is expected, 
we limit the region of interest to within the Rayleigh length $z_R$ 
in order to provide a conservative estimate of the yield. 
Therefore, 
with $V_i = (\pi/2)^{3/2} w^2_{i0} c\tau_i$ from Eq.(\ref{eq_V}),
the overlap factor is approximated as
\beqa
\mcal{D}_I \approx
\Int{-z_R/c}{0}dt
\Int{-\infty}{\infty}dx^i 
\rho_{\mrm{c}}^{2}(t,x^{i})\rho_{\mrm{i}}(t,x^{i})V_{\mrm{i}}
\mbox{\hspace{1.5cm}}  \\ \nnb
= \left( \frac{2}{\pi} \right)^{\frac{3}{2}}\frac{1}{c}\frac{\tau_{\mrm{i}}}{\tau_{\mrm{c}}}
\frac{1}{\sqrt{\tau_{\mrm{c}}^{2}+2\tau_{\mrm{i}}^{2}}}w_{\mrm{i}0}^{2}
\times \mbox{\hspace{3.1cm}}  \\ \nnb
\Int{-z_{R}/c}{0}dt \frac{1}{w_{\mrm{c}}^{4}(ct)+2{w_{\mrm{c}}^{2}(ct)w_{\mrm{i}}^{2}(ct)}},
\mbox{\hspace{2.3cm}}
\eeqa
where the integrand
\beq
\frac{A}{w_{\mrm{c}}^{2}(ct)}-\frac{B}{w_{\mrm{c}}^{2}(ct)+2w_{\mrm{i}}^{2}(ct)} \nnb
\eeq
with
\beq
A  = \frac{1}{2w_{\mrm{i}0}^{2}\left\{ 1-\left( \frac{z_{\mrm{c}R}}{z_{\mrm{i}R}} \right)^{2} \right\}}
\nnb
\eeq
and
\beq
B  = \frac{1}{1-\left( \frac{z_{\mrm{c}R}}{z_{\mrm{i}R}} \right)^{2}}
\left\{ 
\frac{1}{2w_{\mrm{i}0}^{2}}+\frac{1}{w_{\mrm{c}0}^{2}}\left( \frac{z_{\mrm{c}R}}{z_{\mrm{i}R}} \right)^2
\right\}.
\nnb
\eeq
Finally, the spacetime overlap factor for the two effective beams in QPS is expressed as
\beqa\label{eq_Di}
\mcal{D}_I \approx
\sqrt{\frac{2}{\pi}}\frac{1}{c^{2}}\frac{\tau_{\mrm{i}}}{\tau_{\mrm{c}}}\frac{1}{\sqrt{\tau_{\mrm{c}}^{2}+2\tau_{\mrm{i}}^{2}}}
\frac{\lambda_{\mrm{i}}^{2}}{\lambda_{\mrm{i}}^{2}-\lambda_{\mrm{c}}^{2}}
\times \mbox{\hspace{1.0cm}} \\ \nnb
\left[ \frac{1}{\lambda_{\mrm{c}}}\tan^{-1}\left( \frac{\lambda}{\lambda_{\mrm{c}}} \right)
-\frac{1}{\lambda_{\mrm{ci}}}\tan^{-1}\left( \frac{\lambda}{\lambda_{\mrm{ci}}} \right) \right]
\eeqa
with
$\lambda_{\mrm{ci}} \equiv \sqrt{\frac{\lambda_{\mrm{c}}^{2}+2\lambda_{\mrm{i}}^{2}}{3}}$.

\end{document}